\newcommand{\logg}{$\log\,g$}
\newcommand{\teff}{$T_{\rm eff}$}

\documentclass{aa}
\usepackage{txfonts}
\usepackage{graphicx}
\begin{document}

\title{
Wavelengths and oscillator strengths of Xe II from the UVES spectra of four HgMn stars
}

\author{
K.\, Y\"uce
\inst{1,2}
\and
F.\, Castelli
\inst{1}
\and
S.\,Hubrig
\inst{3}
}

\offprints{K. Y\"uce}

\institute{
Istituto Nazionale di Astrofisica,
Osservatorio Astronomico di Trieste, Via Tiepolo 11,
I-34143 Trieste, Italy\\
\email{yuce@oats.inaf.it}
\and
Ankara University, Faculty of Science, Department of Astronomy and Space Sciences,
TR-06100, Tando\u{g}an, Ankara, Turkey
\and
Astrophysical Institute Potsdam, An der Sternwarte 16, D-14482 Potsdam, Germany 
}

\date{}

\abstract
{}
{In spite of large overabundances of \ion{Xe}{ii} observed in 
numerous  mercury-manganese (HgMn) stars, \ion{Xe}{ii} oscillator strengths
are only available for a very limited number of transitions. As a consequence,
several unidentified lines in the spectra of HgMn stars 
could be due to \ion{Xe}{ii}. In addition, some predicted \ion{Xe}{ii} lines 
are  redshifted by about 0.1\,\AA\ from stellar unidentified lines, raising the question
about the wavelength accuracy of  the \ion{Xe}{ii} line data available in the literature.
For these reasons we investigated the \ion{Xe}{ii} lines lying in the 
3900-4521\,\AA, 4769-7542\,\AA, and 7660-8000\,\AA\ spectral ranges of
four well-studied HgMn stars. 
}
{We compared the \ion{Xe}{ii} wavelengths listed
in the NIST database with the position of the lines observed in the 
high-resolution UVES spectrum of the 
xenon-overabundant, slowly
rotating HgMn stars HR\,6000, and we modified them when needed. 
We derived 
astrophysical oscillator strengths for all the \ion{Xe}{ii} observed lines and compared
them with the literature values, when available.
We checked the stellar atomic data derived from HR\,6000 by 
using them to compute synthetic spectra for three other xenon-overabundant, slowly
rotating HgMn stars, HD\,71066, 46\,Aql, and HD\,175640. In this framework, we
performed a complete abundance analysis of HD\,71066, while
we relied on our previous works for the other stars. 
}
{We find that all the lines with wavelengths related to the 6d and 7s energy levels
have a corresponding unidentified spectral line, blueshifted by the same quantity of
about 0.1\,\AA\ in all the four stars, so that we identified these lines as 
coming from \ion{Xe}{ii} and modified their NIST wavelength value according to the observed 
stellar value. We find that the \ion{Xe}{ii} stellar oscillator strengths may differ 
from one star to another from 0.0\,dex  to 0.3\,dex. We adopted 
the average of the oscillator strengths derived from the four stars as final
astrophysical oscillator strength.  
}
{}


\maketitle{}

\section{Introduction}

Several studies of mercury-manganese (HgMn) stars have pointed out 
the presence of xenon with overabundances up to 5\,dex relative to the 
solar value $\log$(N$_{Xe}$/N$_{tot}$)=$-$9.87 (Grevesse \& Sauval 1998).   
This is, for instance,  the case of $\kappa$\,Cnc and 33\,Gem,
for which abundances equal to $-$4.87$\pm$0.13\,dex and 
$-$4.90$\pm$0.07\,dex were determined by Dworetsky et al. (2008).
The xenon overabundance implies the presence of numerous \ion{Xe}{ii} lines 
in the spectra of the HgMn stars, but  the \ion{Xe}{ii} 
transition probabilities are very incomplete, when we compare
the large number of transitions listed in the NIST database and the
small number of them with an associated $\log\,gf$-value. 
As a consequence, the computed spectra do not 
include numerous \ion{Xe}{ii} lines, raising the doubt that some 
unidentified lines could just be due to \ion{Xe}{ii}.
In addition, we noticed that the wavelengths of several \ion{Xe}{ii} lines 
are close, but not coincident with the wavelength 
of some unidentified stellar lines (Castelli \& Hubrig 2007), so that we 
 wondered 
about the accuracy of the wavelength determination from  laboratory spectra.

The most complete work on \ion{Xe}{ii} is that of 
Hansen \& Persson (1987), who analyzed all the 
published (Boyce 1936; Humphreys 1939) and unpublished \ion{Xe}{ii} lines 
from 392\,\AA\ to  10220\,\AA\ obtained in laboratory by Humphreys and Boyce.
In their discussion on the wavelength accuracy, Hansen \& Persson (1987) 
pointed out that the wavelength accuracy for many lines is 
too low to be satisfactory, mostly owing
to the widely varying quality of the experimental data they used.
They announced new experimental work to improve the \ion{Xe}{ii} atomic data.
Unfortunately, this work has never been published up to now, all the more so that
some preliminary results had indicated that, for the high 6d and 7s levels, 
there were
shifts of about 0.5\,cm$^{-1}$ between the energy levels determined 
from the Humphrey wavelengths and the energy levels determined from the new data. This
energy difference  corresponds to a difference of 0.1\,\AA\ in
wavelengths.

Saloman (2004), who performed a critical compilation of all the work 
on energy levels and wavelengths of \ion{Xe}{ii} made up to that time, 
adopted the data from Hansen \& Persson (1987) for almost all 
the lines of the optical region. The Saloman (2004) critical
 compilation is the one adopted by the NIST database. 
 
To study wavelengths and 
$\log\,gf$-values of the \ion{Xe}{ii} lines having intensities
$\ge$ 100 in the NIST line list, 
we used  UVES spectra of the four xenon overabundant HgMn stars 
HR\,6000,  HD\,71066,
46\,Aql, and HD\,175640. They are slowly rotating stars with
v{\it sini}   1.5\,km\,s$^{-1}$, 
1.5\,km\,s$^{-1}$, 1.0\,km\,s$^{-1}$, and 2.5\,km\,s$^{-1}$,
respectively. 

We already performed a complete abundance analysis for HD\,175640 
(Castelli \& Hubrig 2004a)\footnote{http://wwwuser.oat.ts.astro.it/castelli/hd175640/hd175640.html}
and for HR\,6000 and 46\,Aql 
(Castelli et al. 2009)\footnote{http://wwwuser.oat.ts.astro.it/castelli/hr6000new/hr6000.html}.
 To be consistent with the other papers,
we present here an abundance analysis of  HD\,71066, which was studied with the 
same methods as adopted for 
the other stars. A previous work on HD\,71066, related to
vertical abundance stratification in HgMn stars, was performed by 
Thiam et al. (2010), who adopted the same observations as are used
in this paper. We note, however, that no mention about \ion{Xe}{ii} was
made in their study.

\section {Observations  and data reduction}
  
All the stars were observed at the European Southern Observatory (ESO)  
using the Very Large Telescope Ultraviolet and Visible Echelle Spectrograph
(UVES) with a resolving power ranging from 80000 to 110000. 

HD\,175640 was observed on June 13, 2001 (Castelli \& Hubrig 2004a).
HR\,6000, 46\,Aql, and HD\,71066 were
part of the same observational run (ESO program 076.D-0169(A)).
The spectra of HR\,6000  were observed on September 19, 2005,
those of 46 Aql on October 18, 2005 (Castelli et al., 2009), while
the spectra of HD\,71066 were taken on October 27, 2005.
Because Nunez et al. (2010)
found spectral variations in 19 HgMn stars out of a sample of
28 HgMn stars  analyzed, we investigate about a possible variability 
of HD\,71066 by comparing the spectrum observed in 2005 with an UVES
spectrum observed in April 2004. We did not
find any clear indication of variability.

The spectra of the four stars cover the region 3030 $-$ 10000\,\AA.
For HD\,175640 there are  
two gaps at $\lambda\lambda$ 5759 $-$ 5835\,\AA\ and 
8519 $-$ 8656\,\AA. For the other three stars, the gaps occur at 4520 $-$ 4769\,\AA\ and
7536 $-$ 7660\,\AA.
All the spectra were reduced by the UVES pipeline Data Reduction Software
(Ballester et al. 2000).  
We analyzed flux-calibrated spectra for the 3050-5750\,\AA\ region and RED$_{-}$SCI$_{-}$POINT 
spectra for the 5750-9460\,\AA\ interval, in that flux-calibrated reduction
for the red spectra was not implemented in the pipeline reduction procedure.

The measurement procedures on the spectra of HD\,175640, HR\,6000, and
46\,Aql were described in Castelli \& Hubrig (2004a) and 
Castelli \& Hubrig (2007).
The spectra of HD\,71066 were normalized to the continuum using the
IRAF continuum task. 
The equivalent widths were measured by a Gaussian fitting using
the IRAF splot task. 

The S/N ratio is different for the different stars.
In the spectra of HD\,175640,
 it ranges from 200 in the 
near UV to 400 in the visual region. It is higher than the S/N of the
spectra of the other three stars, which were 
observed in a different epoch.
Furthermore, for each star, it is different in the different spectral 
intervals. For instance, for HR\,6000, it is about 100 in the 5800$-$6800\,\AA\
 interval and lowers  to about 25 
at 7400\,\AA\ (REDL spectrum). It is about 50 at 7800\,\AA\ and decreases to 
about 25 at 9400\,\AA\ (REDU spectrum). This behavior is similar for
46 Aql and HD\,71066. At  6800\,\AA\ the S/N is about 100 for HR\,6000, 
70 for 46 Aql,  100 for HD\,71066, and 125 for HD\,175640. 

\begin{table}[]
\begin{center}
\caption{Abundances $\log$(N$_{elem}$/N$_{tot}$) for HD\,71066.}
\begin{tabular}{lccrccccccccccccccc}
\hline\noalign{\smallskip}
\multicolumn{1}{c}{elem}&
\multicolumn{1}{c}{HD\,71066}&
\multicolumn{1}{c}{Star-Sun}&
\multicolumn{1}{c}{Sun$^{a}$}&
\multicolumn{1}{c}{Thiam et al.(2010)}
\\
& [12000K,4.1]& & & [12010,3.95]\\
\hline\noalign{\smallskip}
\ion{He}{i} &$\le$ $-$2.28 &$\le$ [$-$1.23] & $-$1.05 & $-$2.30$\pm$0.40\\
\ion{Be}{ii} &$-$10.79& [$-$0.15] & $-$10.64\\
\ion{C}{ii}  & $-$3.90 &[$-$0.38]  & $-$3.52&$-$3.89$\pm$0.10\\ 
\ion{N}{i}  & $\le$ $-$5.50& $\le$$-$1.38  & $-$4.12\\
\ion{O}{i}  & $-$3.61$\pm$0.05 & [$-$0.40]  & $-$3.21 & $-$3.61$\pm$0.14\\
\ion{Ne}{i}&$\le$ $-$4.70 & $\le$[$-$0.74]        & $-$3.96\\
\ion{Na}{i} & $-$5.51 $\pm$0.08 & [$+$0.20] & $-$5.71\\
\ion{Mg}{i} & $-$5.32 $\pm$0.05 & [$-$0.86]& $-$4.46\\
\ion{Mg}{ii} & $-$5.40 & [$-$0.94]      & $-$4.46 &$-$5.46$\pm$0.01\\
\ion{Al}{i} & $\le$$-$7.30 &$\le$[$-$1.73]   & $-$5.57\\
\ion{Al}{ii} &$\le$$-$7.30 &$\le$[$-$1.73]   & $-$5.57\\
\ion{Si}{ii} & $-$4.61$\pm$0.19 & [$-$0.12]    & $-$4.49&$-$4.58$\pm$0.07\\
\ion{P}{ii} & $-$5.06$\pm$0.13 & [$+$1.53]  & $-$6.59&$-$4.87$\pm$0.22\\
\ion{P}{iii}& $-$5.13          & [$+$1.46]  &$-$6.59\\
\ion{S}{ii} &$-$5.77$\pm$0.11  & [$-$1.06]  & $-$4.71&$-$5.66$\pm$0.20\\
\ion{Ca}{ii} & $-$6.50$\pm$0.21: & [$-$0.82] & $-$5.68&$-$6.02\\
\ion{Sc}{ii} & $\le$$-$$$10.50 &$\le$[$-$1.63]  & $-$8.87\\
\ion{Ti}{ii} & $-$6.45$\pm$0.06 &[$+$0.57]  & $-$7.02 &$-$6.52$\pm$0.05\\
\ion{V}{ii}  &$\le$$-$10.0 & $\le$[$-$1.96]  & $-$8.04\\
\ion{Cr}{ii} & $-$6.17$\pm$0.06 &[$+$0.20] & $-$6.37 &$-$6.28$\pm$0.09\\
\ion{Mn}{ii} & $-$5.95$\pm$0.04 &[$+$0.70] & $-$6.65 & $-$5.81$\pm$0.20\\
\ion{Fe}{i} & $-$3.85$\pm$0.06 & [$+$0.69]     & $-$4.54&$-$3.98$\pm$0.06\\
\ion{Fe}{ii} & $-$3.85$\pm$0.13 &[$+$0.69]  & $-$4.54&$-$3.87$\pm$0.14\\
\ion{Co}{ii} &$\le$$-$7.88 & $\le$[$-$0.76]  & $-$7.12\\
\ion{Ni}{ii} &$\le$$-$7.90 & $\le$[$-$2.11]  & $-$5.79\\
\ion{Cu}{ii} &$\le$$-$7.83 &$\le$[0.00]  & $-$7.83\\
\ion{Zn}{ii} & $\le$$-$7.94&$\le$[$-$0.5]     & $-$7.44\\
\ion{As}{ii} &$-$6.3: &$+$3.37:  & $-$9.67 \\
\ion{Sr}{ii} & $-$8.27 & [$+$0.8] & $-$9.07 &$-$8.35\\
\ion{Y}{ii}  &$-$7.57$\pm$0.08 & [$+$2.23]  & $-$9.80\\
\ion{Xe}{ii} &$-$5.43$\pm$0.16 & [$+$4.44]  & $-$9.87\\
\ion{Nd}{iii}&$-$9.63$\pm$0.01&[$+$0.91] &$-$10.54\\
\ion{Dy}{iii}&$-$9.90&[$+$1.00]  &$-$10.90\\
\ion{Au}{ii} &$-$7.12$\pm$0.03&[$+$3.91] &$-$11.03\\
\ion{Hg}{i} & $-$6.40  &[$+$4.51]    &$-$10.91& $-$6.38$\pm$0.28\\
\ion{Hg}{ii} & $-$6.40 & [$+$4.51]    &$-$10.91&$-$6.53$\pm$0.33\\
\hline\noalign{\smallskip}
\end{tabular}
\end{center}
$^{a}$ Solar abundances are from Grevesse \& Sauval (1998).\\
\end{table}

\begin{table*}[]
\begin{center}
\caption{The strongest emission lines in HD\,71066, with the atomic data 
and configurations from the Kurucz website (see footnote\,5) }
\begin{tabular}{lccrccccccccccccccc}
\hline\noalign{\smallskip}
\multicolumn{1}{c}{$\lambda$($\AA$)}&
\multicolumn{1}{c}{elem}&
\multicolumn{1}{c}{$\log\,gf$}&
\multicolumn{1}{c}{$\chi_{low}$}&
\multicolumn{1}{c}{J$_{low}$}&
\multicolumn{1}{c}{lower config.}&
\multicolumn{1}{c}{$\chi_{up}$}&
\multicolumn{1}{c}{J$_{up}$}&
\multicolumn{1}{c}{upper config.}&
\multicolumn{1}{c}{Rc obs.}&
\multicolumn{1}{c}{Rc comp.}
\\
\hline\noalign{\smallskip}
5987.384 &\ion{Ti}{ii}& $+$0.649 & 64979.278& 3.5 &($^{3}$F)4d e4G & 81676.439 & 4.5& ($^{3}$F)4f 2[4]& 1.012 & 0.983\\
6001.400 &\ion{Ti}{ii}& $+$0.724 & 65095.972& 4.5 &($^{3}$F)4d e4G & 81754.137 & 5.5& ($^{3}$F)4f 2[5]& 1.012 & 0.981\\
6029.278 &\ion{Ti}{ii}& $+$0.653 & 65308.434& 4.5 &($^{3}$F)4d e4H & 81889.576 & 5.5& ($^{3}$F)4f 3[6]& 1.025 & 0.984\\
6125.861 &\ion{Mn}{ii}& $+$0.788 & 82144.480& 3.0 &($^{6}$S)4d e5D & 98464.200 & 4.0& ($^{6}$S)4f $^{5}$F  & 1.023 & 0.896\\
6181.354 &\ion{Cr}{ii}& $+$0.184 & 89812.420& 2.5 &($^{5}$D)4d f4D &105985.630 & 3.5& ($^{5}$D)4f 4[4]& 1.010 & 0.996\\
6182.340 &\ion{Cr}{ii}& $+$0.402 & 89336.890& 2.5 &($^{5}$D)4d e4P &105507.520 & 3.5& ($^{5}$D)4f 2[3]& 1.015 & 0.992\\
6285.601 &\ion{Cr}{ii}& $-$0.229 & 89885.080& 3.5 &($^{5}$D)4d f4D &105790.060 & 4.5& ($^{5}$D)4f $^{4}$F  & 1.011 & 0.998\\
6526.302 &\ion{Cr}{ii}& $+$0.253 & 89885.080& 3.5 &($^{5}$D)4d f4D &105203.460 & 4.5& ($^{3}$F)sp r$^{4}$F & 1.010 & 0.996\\
6551.373 &\ion{Cr}{ii}& $+$0.201 & 90725.870& 3.5 &($^{5}$D)4d e4F &105985.630 & 3.5& ($^{5}$D)4f 4[4]& 1.018 & 0.997\\
6585.241 &\ion{Cr}{ii}& $+$0.815 & 90850.960& 4.5 &($^{5}$D)4d e4F &106032.240 & 5.5& ($^{5}$D)4f 4[6]& 1.028 & 0.987\\
6592.341 &\ion{Cr}{ii}& $+$0.287 & 90512.560& 1.5 &($^{5}$D)4d e4F &105677.490 & 2.5& ($^{5}$D)4f 3[3]& 1.014 & 0.996\\
6636.427 &\ion{Cr}{ii}& $+$0.573 & 90725.870& 3.5 &($^{5}$D)4d e4F &105790.060 & 4.5& ($^{5}$D)4f $^{4}$F  & 1.020 & 0.992\\
6961.439 &\ion{Ti}{ii}& $+$0.663 & 67822.582& 4.5 &($^{3}$F)4d e2G & 82183.467 & 5.5& ($^{3}$F)4f 4[6]& 1.025 & 0.991\\
6982.307 &\ion{Ti}{ii}& $+$0.401 & 67606.162& 3.5 &($^{3}$F)4d e2G & 81924.126 & 4.5& ($^{3}$F)4f 3[4]& 1.015 & 0.995\\
8335.148 &\ion{C}{i}  & $-$0.437 & 61981.820& 1.0 & p3s $^{1}$P    & 73975.910 & 0.0&  p3p $^{1}$S    & 1.023 & 0.889\\                  
9405.730 &\ion{C}{i}  & $+$0.285 & 61981.820& 1.0 & r3s $^{1}$P    & 72610.720 & 2.0&  p3p $^{1}$D    & 1.088 & 0.730 \\ 
\hline\noalign{\smallskip}
\end{tabular}
\end{center}
\end{table*}

\section{The HgMn star HD\,71066}

Previous studies of HD\,71066 ($\kappa^2$ Vol, HR\,3302) have pointed out
the isotopic anomaly of Hg (Dolk et al. 2003; Thiam et al. 2010).
No vertical abundance stratification for Ti, Cr, and Fe is found
by Thiam et al. (2010). No presence of magnetic field is found both 
from the inspection of the
equivalent widths of the \ion{Fe}{ii} lines at
6147.7\,\AA\ and 6149.2\,\AA\ 
( Hubrig et al. 1999) and after using the FORS\,1 spectropolarimeter 
at the VLT (Hubrig et al. 2006).     

\subsection{Model parameters and abundances of HD\,71066}

The starting model parameters of HD\,71066, \teff=12045\,K, and \logg=3.9
were derived both from the Str\"omgren photometry and the 
\ion{Fe}{i} $-$ \ion{Fe}{ii} ionization equilibrium constraint.

The observed colors 
(b$-$y)=$-$0.053, m=0.122, c=0.731 $\beta$=2.769 were taken
from the Hauck \& Mermilliod (1998) Catalogue\footnote{http://obswww.unige.ch/
gcpd/gcpd.html}.
The synthetic colors were taken from the grid computed by Castelli
for [M/H]=0 and microturbulent velocity $\xi$=0\,km\,sec$^{-1}$
\footnote{http://wwwuser.oat.ts.astro.it/castelli/colors/uvbybeta.html}.
Zero reddening was adopted for this star, in agreement with
the results from the UVBYLIST code of Moon (1985).
Observed c and $\beta$ indices are reproduced by
synthetic  indices for model parameters
\teff=12045\,K and \logg=3.9  

\begin{table}[]
\begin{center}
\caption{The isotopic mixture (IM) (in \%) of Hg in HD\,71066 
from the \ion{Hg}{ii} line at 3984\,\AA\ 
as derived by us , Dolk et al. (2003) (DWH), 
and Thiam et al.(2010) (TLKW)}
\begin{tabular}{lcrlrrccccccccccccc}
\hline\noalign{\smallskip}
\multicolumn{1}{c}{isotope}&
\multicolumn{1}{c}{$\lambda$($\AA$)}&
\multicolumn{1}{c}{IM}&
\multicolumn{1}{c}{$\log$(IM)}&
\multicolumn{1}{c}{IM}&
\multicolumn{1}{c}{IM}\\
\multicolumn{1}{c}{}&\multicolumn{1}{c}{} &\multicolumn{2}{c}{this work}& DWH& TLK&\\
\hline\noalign{\smallskip}
196 &3983.771& 0.5 & $-$2.30&0.1$\pm$0.1 & 1.1\\
198 &3983.839& 0.5 & $-$2.30&0.1$\pm$0.2 & 4.0\\
199a&3983.844& 0.5 & $-$2.30&0.1$\pm$0.2 & 3.4\\ 
199b&3983.853& 0.5 & $-$2.30&0.1$\pm$0.2 & 3.4\\
200 &3983.912& 0.5 & $-$2.30&0.1$\pm$0.1 &15.2\\
201a&3983.932& 0.5 & $-$2.30&0.1$\pm$0.3 &66.9\\
201b&3983.941& 0.5 & $-$2.30&0.1$\pm$0.3 &66.9\\
202 &3983.993& 2.5 & $-$1.60&1.5$\pm$0.3 &8.1\\
204 &3984.072& 95.0 & $-$0.022&98.0$\pm$1.5&1.3\\
\hline\noalign{\smallskip}
\end{tabular}
\end{center}
\end{table}

The parameters from the photometry were used for computing an ATLAS9 model
with solar abundances for all the elements 
and zero microturbulent velocity. Using the WIDTH code
(Kurucz 2005), we derived the \ion{Fe}{i} and \ion{Fe}{ii}
abundance from the equivalent widths of 12 \ion{Fe}{i} lines
and 26 \ion{Fe}{ii} lines. Seven of the \ion{Fe}{ii} lines are
transitions between high-excitation energy levels, and they 
have experimental $\log\,gf$-values. They
were used to determine the iron abundance, in that they are
rather independent of \teff\ and \logg\ (Castelli et al., 2009). 
Then, we searched for the model atmosphere giving this same abundance
from both
\ion{Fe}{i} lines and low-excitation \ion{Fe}{ii} lines. 
All the adopted lines are listed in Table\,A.1 of Appendix\,A (online material).
We find that the ATLAS9 
model with the parameters \teff=12045\,K, \logg=3.9 derived
from the Str\"omgren photometry meets
the requirement of same iron abundance from all the different
kinds of iron lines. In fact, it gives an average
abundance $\log$(N(\ion{Fe})/N$_{tot}$) equal to 
$-$3.88$\pm$0.08 from the \ion{Fe}{i} lines,
$-$3.92$\pm$0.12 from the low-excitation \ion{Fe}{ii} lines,
and $-$3.84$\pm$0.05 from the \ion{Fe}{ii} high-excitation lines.

The ATLAS9 model was used to derive the abundance for all
those  elements that show lines in
the synthetic spectrum when  solar abundance is adopted
for them.
Whenever possible,  equivalent widths were measured to derive
the abundances. For weak and blended lines and for lines that are blends of 
transitions belonging to the same multiplet, such as
\ion{Mg}{ii} 4481\,\AA, \ion{He}{i} lines, and most \ion{O}{i}
profiles, we derived the abundance from the line profiles.
The synthetic spectrum was also used to determine upper
abundance limits from those lines predicted for solar
abundances, but not observed.

\begin{table*}[]
\caption[ ]{Atomic data for selected Xe II lines} 
\begin{flushleft}
\begin{tabular}{rrrllrrcllllll}
\hline\noalign{\smallskip}
\multicolumn{1}{c}{$\lambda$(Ritz)}&
\multicolumn{1}{c}{Int.}&
\multicolumn{1}{c}{$\chi_{low}$}&
\multicolumn{2}{c}{Term}&
\multicolumn{1}{c}{$\chi_{up}$}&
\multicolumn{2}{c}{Term}&
\multicolumn{1}{c}{$\log~gf$}&
\multicolumn{2}{c}{$log(\gamma_{S}$/Ne)(cm$^{3}$\,s$^{-1}$)}\\
\hline\noalign{\smallskip}
\multicolumn{1}{c}{($\AA$)}&
\multicolumn{1}{c}{}&
\multicolumn{1}{c}{(cm$^{-1}$)}&
\multicolumn{2}{c}{}&
\multicolumn{1}{c}{(cm$^{-1}$)}&
\multicolumn{2}{c}{}&
\multicolumn{1}{c}{NIST}&
\multicolumn{1}{c}{Dj}&
\multicolumn{1}{c}{PD}\\
\hline\noalign{\smallskip}
    4844.32& 2000     &  93068.44 &($^{3}$P$_{2}$)6s & [2]$_{5/2}$&113705.40& ($^{3}$P$_{2}$)6p&[3]$_{7/2}$&  $+$0.491  &$-$5.347&$-$5.420\\
    5292.21& 1000     &  93068.44 &($^{3}$P$_{2}$)6s & [2]$_{5/2}$&111958.89& ($^{3}$P$_{2}$)6p&[2]$_{5/2}$&  $+$0.351  &$-$5.482 &$-$5.450\\
    5419.14& 2000     &  95064.38 &($^{3}$P$_{2}$)6s & [2]$_{3/2}$&113512.36& ($^{3}$P$_{2}$)6p&[3]$_{5/2}$&  $+$0.215  &$-$5.481&$-$5.518 \\
    5438.97&  400     & 102799.07 &($^{3}$P$_{1}$)6s & [1]$_{3/2}$&121179.80& ($^{3}$P$_{1}$)6p&[0]$_{1/2}$&  $-$0.183  &$-$5.544&$-$5.369\\
    5472.61&  500     &  95437.67 &($^{3}$P$_{2}$)5d & [3]$_{7/2}$&113705.40& ($^{3}$P$_{2}$)6p&[3]$_{7/2}$&  $-$0.449  &$-$5.482\\
    5531.06&  400     &  95437.67 &($^{3}$P$_{2}$)5d & [3]$_{7/2}$&113512.36& ($^{3}$P$_{2}$)6p&[3]$_{5/2}$&  $-$0.616  &$-$5.504\\
    5719.61&  200     &  96033.48 &($^{3}$P$_{2}$)5d & [2]$_{3/2}$&113512.36& ($^{3}$P$_{2}$)6p&[3]$_{5/2}$&  $-$0.746  &\\
    5976.46& 1000     &  95064.38 &($^{3}$P$_{2}$)6s & [2]$_{3/2}$&111792.17& ($^{3}$P$_{2}$)6p&[2]$_{3/2}$&  $-$0.222  & $-$5.545&$-$5.556\\
    6036.20&  500     &   95396.74 &($^{3}$P$_{2}$)5d & [2]$_{5/2}$&111958.89& ($^{3}$P$_{2}$)6p&[2]$_{5/2}$& $-$0.609 &$-$5.535 &  \\
    6051.15 &1000     &   95437.67 &($^{3}$P$_{2}$)5d & [3]$_{7/2}$&111958.89& ($^{3}$P$_{2}$)6p&[2]$_{5/2}$& $-$0.252  &$-$5.515 &  \\
    6097.59 &1000     &   95396.74 &($^{3}$P$_{2}$)5d & [2]$_{5/2}$&111792.17& ($^{3}$P$_{2}$)6p&[2]$_{3/2}$& $-$0.237  &\\
    6990.88 &2000     &   99404.99 &($^{3}$P$_{2}$)5d & [4]$_{9/2}$&113705.40& ($^{3}$P$_{2}$)6p&[3]$_{7/2}$& $+$0.200  &  &$-$5.476\\
\hline\noalign{\smallskip}
\end{tabular}
\end{flushleft}
\end{table*}

\begin{table*}
\begin{flushleft}
\caption{Xenon abundance from the measured equivalent widths of HR\,6000, 46\,Aql,
HD\,71066, and HD\,175640, for each star, using ATLAS12 models with parameters \teff\ and
\logg\ given in the table.} 
\begin{tabular}{l|rl|rl|rl|rll}
\noalign{\smallskip}\hline
\multicolumn{1}{c}{}&
\multicolumn{2}{c}{HR\,6000}&
\multicolumn{2}{c}{HD\,71066}&
\multicolumn{2}{c}{46 Aql}&
\multicolumn{2}{c}{HD\,175640}
\\
\multicolumn{1}{c}{[\teff,\logg]}&
\multicolumn{2}{c}{[13450,4.40]}&
\multicolumn{2}{c}{[12000,4.10]}&
\multicolumn{2}{c}{[12560,3.80]}&
\multicolumn{2}{c}{[12000,3.95]}
\\
\noalign{\smallskip}\hline
\multicolumn{1}{c}{$\lambda$($\AA$)}&
\multicolumn{1}{c}{W(m\AA)}&
\multicolumn{1}{c}{abund}&
\multicolumn{1}{c}{W(m\AA)}&
\multicolumn{1}{c}{abund}&
\multicolumn{1}{c}{W(m\AA)}&
\multicolumn{1}{c}{abund}&
\multicolumn{1}{c}{W(m\AA)}&
\multicolumn{1}{c}{abund}\\
\noalign{\smallskip}\hline
4844.33  &28.80 & $-$5.10 & 20.72& $-$5.21  &17.94  &$-$5.67  & 11.33 & $-$5.86\\
5292.21  &30.19 & $-$4.98 &  19.72 & $-$5.20& 16.52  &$-$5.63  &10.80 & $-$5.81\\
5419.14  &23.63 & $-$5.02 &  14.27 & $-$5.24 & 13.34  &$-$5.56  &7.75 & $-$5.80\\
5438.97  &5.71  & $-$5.42  &   2.93 & $-$5.55& 2.59  &$-$5.85  & $--$ & $--$\\
5472.61  &7.56  & $-$5.34  &   5.03 & $-$5.34& 2.85  &$-$5.90  & $--$ & $--$\\
5531.06  &4.13  & $-$5.52  &   1.87 & $-$5.71& 2.05  &$-$5.89  & $--$ & $--$\\
5719.61  &4.29  & $-$5.30  &   1.40 & $-$5.64 &$--$  &$--$     & 1.79 & $-$5.58\\
5976.46  &11.34 &$-$5.18  &   4.74 & $-$5.49 &3.43  &$-$5.93   & 1.60 & $-$6.15\\
6036.20  &6.74  &$-$5.14  &   2.44 & $-$5.45  & 2.31  & $-$5.73 & $--$& $--$\\
6051.15  &9.56  &$-$5.25  &  4.59 & $-$5.44  &3.65  &$-$5.84  &  1.46 & $-$6.13\\
6097.59  &6.79  &$-$5.49  &  3.93 &$-$5.53   &2.57  &$-$6.03   & 1.94 & $-$5.99\\
6990.88  & 11.20 &$-$5.18  &   5.18 & $-$5.36 &5.59  &$-$5.61  & 2.52 & $-$5.86\\
\noalign{\smallskip}\hline
aver abund. &       \multicolumn{2}{c}{$-$5.25$\pm$0.17}  &\multicolumn{2}{c}{$-$5.43$\pm$0.16}&\multicolumn{2}{c}{$-$5.79$\pm$0.15}&
\multicolumn{2}{c}{$-$5.90$\pm$0.17}\\
\noalign{\smallskip}\hline
\end{tabular}
\end{flushleft}
\end{table*}

 The SYNTHE code, together with updated Kurucz line lists
(Castelli \& Hubrig 2004a; Castelli \& Kurucz 2010), were
used to compute the synthetic spectrum. 
The synthetic spectrum was broadened for the instrumental profile
and for a rotational velocity {\it vsini}=1.5\,km\,s$^{-1}$, which
was derived from the comparison of the observed and computed
profile of \ion{Mg}{ii} at 4481\,\AA.

Once all the abundances had been determined in this way, we computed an ATLAS12
model for the individual abundances having the same parameters as
the ATLAS9 model. 
We used the seven \ion{Fe}{ii} high-excitation lines to determine the
new iron abundance. The ATLAS12 parameters were then modified until obtaining 
the same iron abundance, within the error limits, from both the
\ion{Fe}{i} lines and the low-excitation \ion{Fe}{ii} lines. 
The average iron abundances from \ion{Fe}{i}, \ion{Fe}{ii} low-excitation, and  
\ion{Fe}{ii} high-excitation lines are 
 $-$3.85$\pm$0.07, $-$3.87$\pm$0.12, and $-$3.81$\pm$0.05,
respectively, for an  ATLAS12 model with parameters 
\teff=12000\,K and \logg=4.1. This model also leads to
good agreement between the observed and computed
 H$_{\alpha}$ profiles. 
The abundances of HD\,71066 derived from the ATLAS12 model 
either from equivalent
widths or line profiles are listed in Table\,1.

We also see \ion{As}{ii} lines at
4466.348 (weak), 4494.23 (weak),
5105.58, 5231.38, 5331.23, 5497.727, 5558.09, 5651.32, 6110.07,
and 6170.27\,\AA\ were observed in the spectrum.
 Owing to the lack of $\log\,gf$-values for all the optical \ion{As}{ii}
transitions, we can only infer  overabundance of this element in 
HD\,71066. A guessed abundance of $-$6.3 dex for arsenic was derived from  
measured equivalent widths and from guessed $\log\,gf$-values
(Table\,A.1 and Table\,1).

In addition to the overabundance of arsenic and to the large overabundance 
of iron ([$+$0.69]),
overabundances of P ([$+$1.5]), Na ([$+$0.2]), Ti ([$+$0.6]), 
Cr ([$+$0.2]), Mn ([$+$0.7]), Sr ([$+$0.8]), Y([+2.2]), Xe ([+4.4]),
Nd([+0.9]), Dy ([+1.0], Au ([+3.9]), and Hg ([$+$4.5]) were observed.
The other elements - He, Be, C ,N, O, Ne, Mg, Al, Si, S, Ca, S,
V, Co, Ni, Cu, and Zn - are underabundant.

No \ion{Pt}{ii} lines were observed. A weak line at
4046.58\,\AA\ is \ion{Hg}{i} at 4046.56, which is surely not blended
with \ion{Pt}{ii} at 4046.433\,\AA, because the spectral resolution
is high enough, and the rotational velocity is low enough to permit
us to see \ion{Pt}{ii} when it is present. Furthermore, the line
observed is reproduced well by assuming the mercury abundance and
the isotopic composition deduced from \ion{Hg}{ii} at 3984\,\AA\ (Sect. 3.3).    

The comparison with the abundances by Thiam et al. (2010) has shown close
agreement  between the two determinations. Because Thiam et al. (2010) use
an ATLAS9 model computed for solar abundances, the ATLAS12 model
computed for an individual abundance may be estimated as unnecessary. 
However, in addition to the closer
values for the \ion{Fe}{i} and \ion{Fe}{ii} abundances obtained with the 
ATLAS12 model, the
consistency of the elemental abundances in the model and in the synthetic
spectrum generally gives better agreement between the observed and computed
profiles, in particular for the hydrogen profiles when He-weak stars are concerned.

\subsection{Emission lines}

Emission lines were observed for \ion{C}{i}, \ion{Ti}{ii}, \ion{Cr}{ii},
\ion{Mn}{ii}, and possibly for \ion{Fe}{ii}.
Most emissions are so weak that we stated their
presence mostly on the basis of the emissions observed in other stars,
in particular HD\,175640 (Castelli \& Hubrig 2004a).
The emissions greater than the spectral noise are
those listed in Table\,2. The atomic data are taken from the Kurucz
database\footnote{http://kurucz.harvard.edu/atoms.html}.
For \ion{Mn}{ii} mult.\,13, only the transition at 6125.861\,\AA\ shows
true emission, while the other lines at $\lambda\lambda$
6122.434, 6126.225, 6128.734, 6129.033, 6130.796, and
6131.923\,\AA\ are observed to be much weaker than computed,
so that we assume that also these Mn II lines are affected by emission.

\begin{figure}
\centering
\resizebox{4.50in}{!}{\rotatebox{90}{\includegraphics[0,0][400,800]
{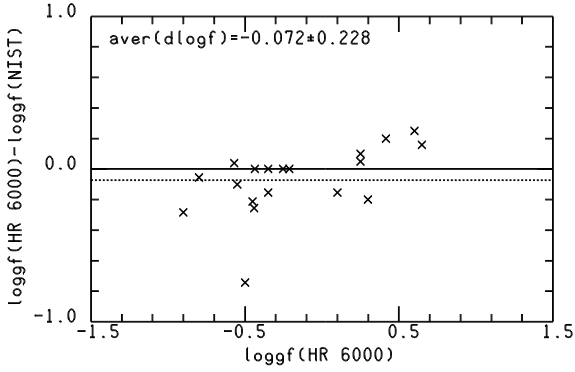}}}
\vskip -0.2cm
\caption{Comparison of astrophysical $\log\,gf$-values of \ion{Xe}{ii} derived
from HR\,6000 with the $\log\,gf$-values of the NIST critical compilation. }
\end{figure}

\begin{figure}
\centering
\resizebox{4.50in}{!}{\rotatebox{90}{\includegraphics[0,0][400,800]
{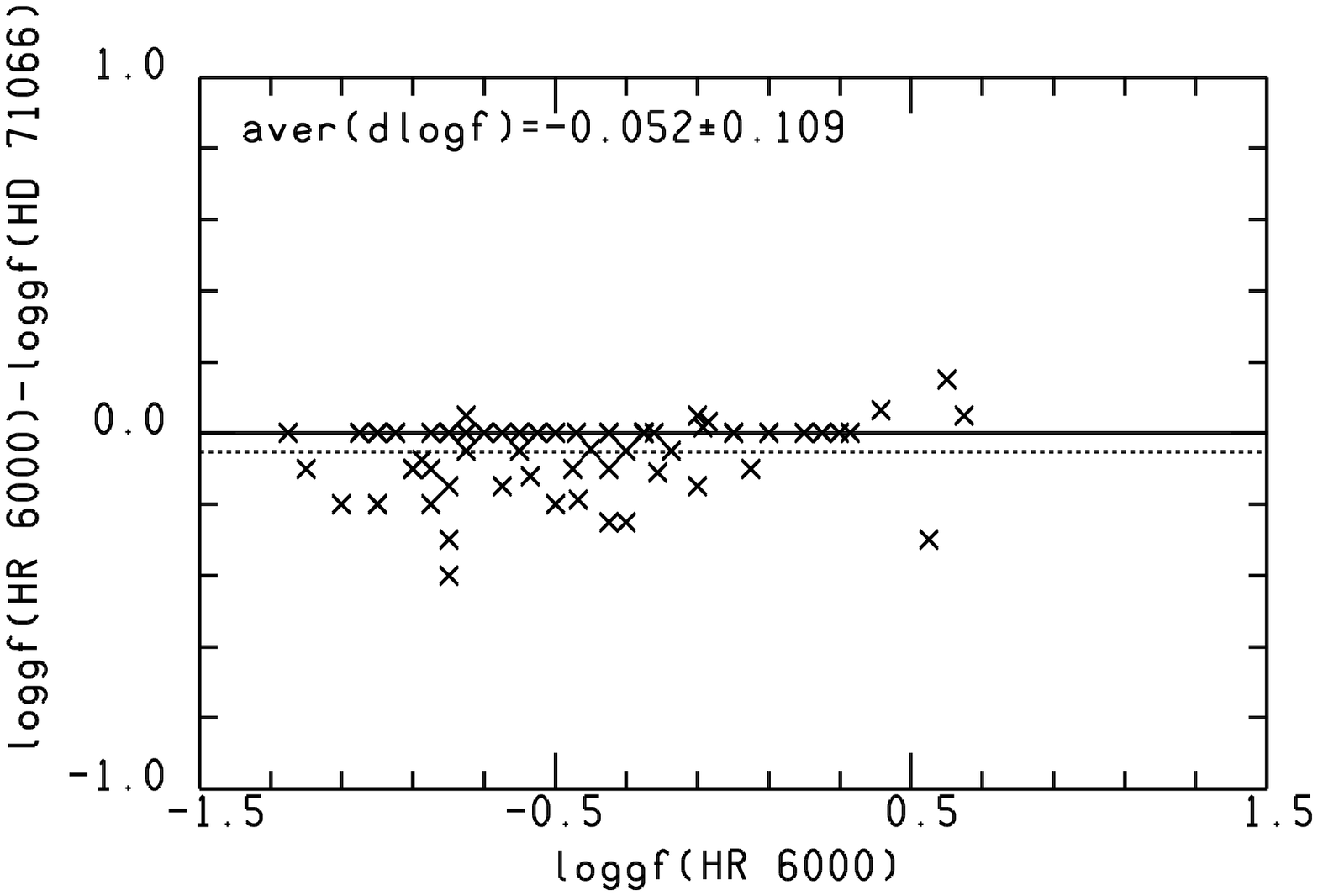}}}
\vskip -0.2cm
\caption{Comparison of astrophysical $\log\,gf$-values of \ion{Xe}{ii} derived
from HR\,6000 with the astrophysical $\log\,gf$-values derived from HD\,71066. }
\end{figure}

\begin{figure}
\centering
\resizebox{4.50in}{!}{\rotatebox{90}{\includegraphics[0,0][450,800]
{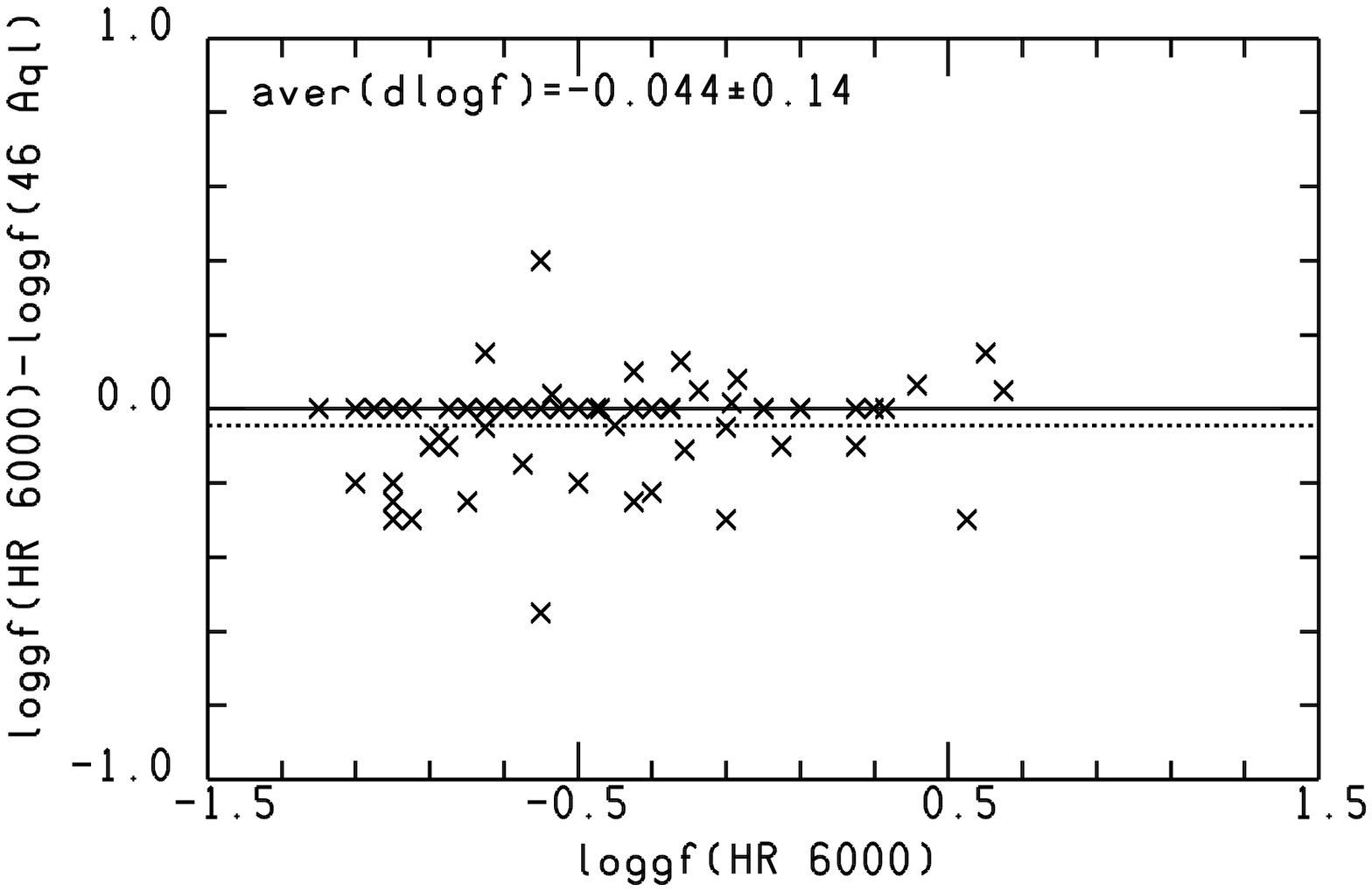}}}
\vskip -0.2cm
\caption{Comparison of astrophysical $\log\,gf$-values of \ion{Xe}{ii} derived
from HR\,6000 with the astrophysical $\log\,gf$-values derived from 46\,Aql. }

\end{figure}

\begin{figure}
\centering
\resizebox{4.50in}{!}{\rotatebox{90}{\includegraphics[0,0][400,800]
{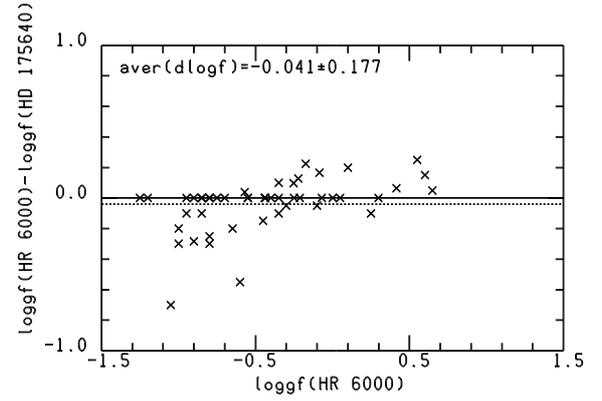}}}
\vskip -0.2cm
\caption{Comparison of astrophysical $\log\,gf$-values of \ion{Xe}{ii} derived
from HR\,6000 with the astrophysical $\log\,gf$-values derived from HD\,175640.}

\end{figure}

\subsection{Isotopic anomalies}

We found an anomalous isotopic composition in HD\,71066
for mercury and caicium. Dolk et al. (2003) have determined an anomalous isotopic composition
for Hg by analyzing the line of \ion{Hg}{ii} at 3984\,\AA.
Table\,3 shows that our results agree with theirs, 
while they are somewhat different from those of Thiam et al. (2010).
We also obtained very good agreement between the observed and
computed line \ion{Hg}{i} at 4046.5\,\AA\ by adopting the
same isotopic composition and abundance ($-$6.4\,dex)
derived from \ion{Hg}{ii} at 3984\,\AA.

The lines of the \ion{Ca}{ii} infrared triplet at $\lambda\lambda$
8498.023, 8542.091, and 8662.14\,\AA\ are redshifted
by 0.16\,dex. Such a shift, observed in numerous HgMn stars
 and Ap stars (Cowley et al. 2007) was discovered by 
Castelli \& Hubrig (2004b), who interpret it as due to an anomalous 
calcium isotopic composition.

\section {The Xenon abundance in HR\,6000,  HD\,71066, 46\,Aql, and HD\,175640}

To compute the \ion{Xe}{ii} line spectrum we derived
the xenon abundance in each star from the equivalent widths
of a set of unblended \ion{Xe}{ii} lines. The WIDTH code
(Kurucz 2005) was used.
The selected lines and their atomic data are 
listed in Table\,4.
Wavelengths and $\log\,gf$-values were taken from
the NIST database (Version\,4)\footnote{http://www.nist.gov/pml/data/asd.cfm}. 
We assumed the classical radiative damping
constant $\gamma_{R}$=0.2223x10$^{16}$/$\lambda$$^{2}$ s$^{-1}$ for $\lambda$ in \AA.
For Stark broadening we used the experimental results from Djurovic et al. (2006).
Because they are given for a temperature of T=22000\,K, we investigated the effect
of the temperature on the Stark damping constant $\gamma_{S}$.
We interpolated for T=12000\,K in
the tables from Popovic \& Dimitrijevic (1997),  which list Stark widths computed
at different temperatures. The last two columns
of Table\,4 compare
 $\gamma_{S}$ values from Djurovic et al. (2006) (Dj)  with the
interpolated values for temperature from Popovic \& Dimitrijevic (1997)(PD). 
We found that the differences in $\gamma_{S}$ from the two sources
do not affect the abundances more than 0.01\,dex.
The approximations of the WIDTH code were used (Castelli 2005) f
for no available Stark damping constants and for Van der Waals damping constants. 

The measured equivalent widths of the selected \ion{Xe}{ii} lines and the 
corresponding abundances are given in Table\,5.

\section{Stellar wavelengths and the astrophysical $\log\,gf$-values for Xe II}

Because xenon is more abundant in HR\,6000 than in the other stars,
we searched in the HR\,6000 spectra for those \ion{Xe}{ii} lines with an intensity
equal to or higher than 100 in the NIST line list.
When these lines were observed in the spectra, they were added in our line list. 
For lines with  no  available $\log\,gf$-values, we
assigned  guessed  values based on  the line intensity.
We examined the interval 3900-8000\,\AA\ with two gaps in the 
4525 $-$ 4780\,\AA\ and 7536 $-$ 7660\,\AA\ regions, due to the lack of spectra in
these ranges.

A synthetic spectrum for HR\,6000 was computed for the xenon abundance given in Table\,5
and for the abundances of all the other elements as derived by Castelli et al. (2009).
In all the stars, the wavelength scale was fixed by shifting the observed spectrum
on the computed spectrum until overimposing some lines with well-determined 
wavelength values such as \ion{Ca}{ii} 3933.663\,\AA, \ion{Mg}{ii} 4481.126\,\AA,
4481.150\,\AA\,, 4481.325\,\AA\,, and several strong \ion{Fe}{ii} lines.  

For all the considered \ion{Xe}{ii} lines, we adjusted the $\log\,gf$-value until
 the observed and computed profiles agree best.
For several lines we also adjusted the NIST wavelength, because we noticed that,
while they do not have an observed counterpart, they are close to an unidentified stellar line 
with wavelength blueshifted  up to 0.1\,\AA\ from the predicted \ion{Xe}{ii} line.  

The astrophysical $\log\,gf$-values and the adjusted wavelengths were then checked
on the three other stars by comparing their observed spectra with
synthetic spectra computed with the \ion{Xe}{ii} wavelengths and oscillator
strengths obtained from the spectrum of HR\,6000. The \ion{Xe}{ii} abundances 
adopted for the three stars are those given in Table\,5. 
Table\,B.1 in Appendix\,B lists
wavelengths and $\log\,gf$-values as derived from the four stars.
We found that for all the examined transitions, the stellar wavelength
is the same in the four stars, except for the lines at 5260.44\,\AA\
and 6343.96\,\AA. The largest difference between stellar and NIST
wavelength is $-$0.13\,\AA\ observed for the line at 4330.52\,\AA.
This line, as well as all the other lines with $\Delta\lambda$$\sim$$-$0.1\,\AA\,
has a 6d or a 7s level as upper level. The uncertainty of the energy of these levels
is on the order of 0.5\,cm$^{-1}$ according to Hansen \& Persson (1987). 

\begin{figure*}
\centering
\resizebox{7.50in}{!}{\rotatebox{90}{\includegraphics[50,115][600,765]
{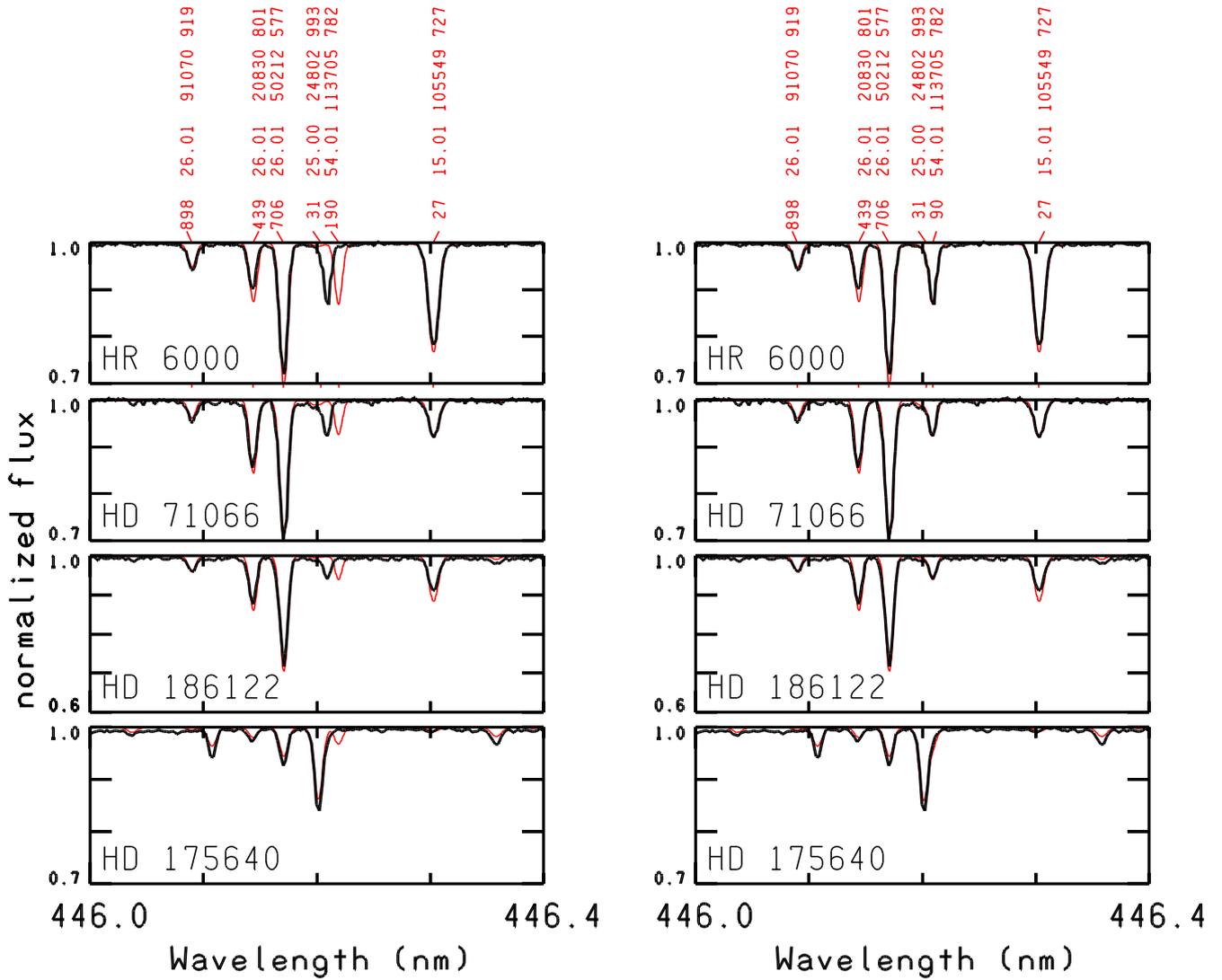}}}
\caption{Comparison for the four stars of the observed (black line) and 
computed (red line)
spectra in the region of the \ion{Xe}{ii} line with wavelength 446.2190\,nm
according to the NIST database (left panel) and with wavelength 446.2090\,nm
according to this paper (right panel). The line identification can be decoded
as follows: for the first line, 898 last 3 digits of wavelength 446.0898\,nm;
26 atomic number of iron; .01 charge/100; i.e., 26.01 identifies the 
line as \ion{Fe}{ii}; 91\,070 is the energy of the lower level in cm$^{-1}$;
919 is the residual central intensity in per mil. The \ion{Xe}{ii} line is 
identified by 54.01.
}

\end{figure*}

Figures 1, 2, 3, and 4 show the comparison of the HR\,6000 astrophysical
$\log\,gf$-values for \ion{Xe}{ii} with the $\log\,gf$-values taken from
the NIST database and with the the astrophysical $\log\,gf$-values 
derived
from the spectra of HD\,71066, 46\,Aql, and HD\,175640, respectively (Table\,B.1, col.\,5).
The largest discrepancy with the NIST data occurs for the line
at 4414.84\,\AA. We adopted the stellar $\log\,gf$-value for it because
it gives an excellent agreement between   
the observed and computed profiles in all the four stars we examined.
The comparison of the astrophysical $\log\,gf$-values of HR\,6000 with those
from the other stars shows that they are on average
lower by about 0.04-0.05\,dex than those from the other stars and that the mean 
square deviation from
the average increases with the decrease in the xenon abundance.
We note that the weaker a line, the more uncertain 
its astrophysical $\log\,gf$-value, mostly when the noise is not
negligible. In particular, red spectra are affected 
both by rather large noise and by numerous telluric lines that lower
the accuracy of the results.

The final line list for \ion{Xe}{ii} is shown in Table\,6. Columns
1 and 2 give the wavelength derived from the stellar spectra and
the astrophysical $\log\,gf$-value obtained by averaging the astrophysical
$\log\,gf$-values from the four stars. The associated error is
the standard deviation from the mean. When it is not given, it means that
the $\log\,gf$-value was obtained from only one star. Columns 3 and 4
list $\log\,gf$-values from the literature and the source. The
literature sources are  
the NIST database, version 4 (NIST4), and  Z\'ieli\'nska et al. (2002) (ZBD).
Z\'ieli\'nska et al. (2002) estimate that, in general, their experimental
transition rates agree with the NIST critical
compilation made by Reader et al. (1980), which is the one adopted in
the NIST4 database. 

The last column gives the  $\gamma_{Stark}$ parameter, which was determined as 
described in Sect.\,4. Figure\,5 shows, for each studied star, the comparison of the observed and
computed spectra in the region of the \ion{Xe}{ii} line with wavelength
4462.190\,\AA, according to the NIST database, and 4462.090\,\AA, according
to Table\,6. The wavelength shift of 0.1\,\AA\ 
and the astrophysical $\log\,gf$ value of $+$0.33, which are the same for all the
stars, provide excellent agreement between the observed and computed
\ion{Xe}{ii} profiles.

\section{Conclusions}

From the high resolution stellar spectra of four HgMn stars we derived 
both wavelengths and $\log\,gf$-values for
100 \ion{Xe}{ii} lines, which should also be observable in the spectra of 
numerous others 
chemically peculiar B-type stars. Of these lines, only 22 lines have
$\log\,gf$-values available in the NIST database. The NIST wavelength of two 
of them, 4180.10\,\AA\ and 4330.52\,\AA,
differs by about 0.1\,\AA\ from that observed in the spectra. There is
a total of 27 lines in our sample for which the observed wavelength
differs from the NIST wavelength by more than $-$0.06\,\AA\ with
the maximum shift of $-$0.13\,\AA\  for the line 
at 4330.52\,\AA. We believe that the wavelength differences are mostly
the result of uncorrect energy levels, in that they are all related to 6d or 7s
levels, which have an uncertainty of about 0.5\,cm$^{-1}$ according to
Hansen \& Persson (1987). This hypothesis seems us to be more relastic 
than that of some isotopic anomaly for \ion{Xe}. 
For instance, using the isotopic wavelengths
from Alvarez et al. (1979), Castelli \& Hubrig (2007) excluded that 
the blueshift  of 0.03\,\AA\ observed for the \ion{Xe}{ii} line at 
6051.15\,\AA\ can be due to some isotopic anomaly.
Instead, because no isotopic composition was considered in our
computations, owing to the lack of isotopic wavelengths for 
\ion{Xe}{ii}, we could explain the larger astrophysical $\log\,gf$-value
than the experimental one obtained for a few lines
with the presence of the xenon isotopes,
which should not be neglected in the computations of the strongest \ion{Xe}{ii}
line profiles. Good examples are  the lines at 4844.33\,\AA,   
5292.22\,\AA, and  5419.155\,\AA\ (Table\,6).  

On the basis of the wavelength shifts observed in the stellar spectra we
redetermined the energy of three 7s, one 5d, and eighteen 6d levels.
These levels, together with the old and new energy values, are listed in
Table\,7. We would like to point out that the new energy values 
depend, of course,  on the accuracy of the energy of the lower level.

The identification of the \ion{Xe}{ii}
lines and their consequent addition in the line lists, increases the accuracy of 
the synthetic spectra for the CP stars. In fact, it is important to
be able to reproduce their high-resolution spectra well, because 
these stars are an excellent tool for extending laboratory spectrum
analyses for several elements. An example is the determination of new
high-excitation energy levels for \ion{Fe}{ii} from the same UVES spectra
of HR\,6000 used for this paper (Castelli \& Kurucz 2010).
For instance, \ion{As}{ii} is another element observed in some CP stars for
which not even one $\log\,gf$-value in the optical region has been found in the literature.
\ion{As}{ii} has not only been observed in 46\,Aql 
(Sadakane et al. 2001, Castelli et al. 2009), but also in HD\,71066,
as we have shown in this paper.
If only one $\log\,gf$ value were given for it, we could derive
astrophysical $\log\,gf$-values for the other lines, just as we did for
\ion{Xe}{ii}. 

The abundance analysis of HD\,71066 has pointed out the overabundances of \ion{Y}{ii},
\ion{Nd}{iii}, \ion{Dy}{iii}, and \ion{Au}{ii} for the first time,  
in addition to the \ion{Xe}{ii} and \ion{As}{ii} overabundances. Those
of  other elements, in particular Hg, P, Ti, Cr, Mn, Fe, and Sr,
have already been stated by Thiam et al. (2010) and confirmed by us.

We found that HD\,71066 is a typical HgMn star with Hg and Ca isotopic
anomalies and emission lines for \ion{C}{i},  \ion{Ti}{ii}, \ion{Cr}{ii},
and  \ion{Mn}{ii}. \ion{He}{i} is underabundant and  
the shape of its profiles indicates the presence of helium vertical
abundance stratification in the atmosphere.

\begin{table}[!hbp]
\caption[ ]{The final \ion{Xe}{ii} astrophysical line list for 
the 3900-4525\,\AA\ and 4780-8000\,\AA\ intervals. The literature 
$\log\,gf$ sources are the NIST database, version 4 (NIST4) and
 Z\'ieli\'nska et al. (2002)(ZBD).}
\font\grande=cmr7
\grande
\begin{flushleft}
\begin{tabular}{lllllrrcllllll}
\hline\noalign{\smallskip}
\multicolumn{1}{c}{$\lambda$}&
\multicolumn{1}{c}{$\log~gf$}&
\multicolumn{1}{c}{$\log~gf$}&
\multicolumn{1}{c}{source}&
\multicolumn{1}{c}{$log\gamma_{S}$}\\
\hline\noalign{\smallskip}
\multicolumn{1}{c}{stellar}&
\multicolumn{1}{c}{stellar}&
\multicolumn{1}{c}{literature}& & &\\
\hline\noalign{\smallskip}
  3907.820 & $-$0.82$\pm$0.06 &    &     &$-$4.684&\\
  4037.260 & $-$1.00$\pm$0.00 &    &       \\
  4037.470 & $-$0.75$\pm$0.00 & \\
  4057.360 & $-$0.80$\pm$0.00 &    &     &$-$4.899 \\
  4157.980 & $-$0.60$\pm$0.00 &    &     &$-$4.878 \\
  4162.160 & $-$1.57$\pm$0.03 &  &       &$-$5.379 \\
  4180.007 & $-$0.35$\pm$0.00 & $-$0.35 &NIST4&         \\
  4193.100 & $-$0.60  \\
  4208.391 & $-$0.38$\pm$0.02 \\
  4209.370 & $-$0.70$\pm$0.00\\
  4213.620 & $-$0.22$\pm$0.08&\\
  4215.620 & $-$1.05$\pm$0.00&\\
  4222.900 & $+$0.64$\pm$0.23&   &        &$-$4.778  \\
  4238.135 & $-$0.23$\pm$0.10&   &      &$-$4.948 &\\
  4245.300 & $-$0.13$\pm$0.07&   &      &$-$4.930 &\\
  4251.540 & $-$0.58$\pm$0.02&   &      &$-$4.722 &\\
  4296.320 & $-$0.85$\pm$0.00&     &      &$-$5.129 &\\
  4330.390 & $+$0.30$\pm$0.00 & $+$0.498&NIST4&$-$4.884 &\\
  4369.100 & $-$0.72$\pm$0.02&   &      &$-$4.890 & \\
  4373.700 & $-$0.70$\pm$0.00 \\
  4384.910 &$\le$ $-$1.95 & & &$-$5.358\\
  4393.090 & $+$0.00$\pm$0.00&      &     &$-$4.927 &\\
  4395.770 & $+$0.00$\pm$0.00&      &     &$-$4.884 &\\
  4414.840 & $-$0.50$\pm$0.00& $+$0.243&NIST4&$-$5.432  \\
  4416.090 & $-$0.80& \\
  4448.025 & $+$0.10$\pm$0.05\\
  4462.090 & $+$0.33$\pm$0.00 &     &     &$-$4.866\\
    $----$\\
  4787.77  &$-$0.82$\pm$0.03    &  &    &$-$5.324\\
  4817.98  &$-$1.25$\pm$0.00   &    &    &$-$5.351 \\
  4823.25  &$-$0.65$\pm$0.00   &    &    &$-$4.989\\
 4844.33  &$+$0.61$\pm$0.02& $+$0.491 &NIST4 & $-$5.347\\
         &                   & $+$0.510$\pm$0.027&ZBD\\ 
 4876.50 & $+$0.10$\pm$0.00 & $+$0.255&NIST4 & $-$5.505 \\
 4883.53 & $-$0.25$\pm$0.00 &         &                    &$-$5.525  \\    
 4884.09 & $-$0.80  \\    
 4887.30 & $-$0.85$\pm$0.05   &         &                    &$-$5.423 &\\
 4890.085& $-$1.17$\pm$0.04  &$-$0.754$\pm$0.022&ZBD&$-$5.420 & \\
 4919.66 & $-$0.85$\pm$0.12 & \\ 
 4921.48 & $+$0.05$\pm$0.09&  &  &$-$4.442 &  \\
 4971.68 & $-$0.75$\pm$0.00 \\
 4972.70 & $-$0.55$\pm$0.00   &  &  &$-$5.430 & \\
 4988.725& $-$0.85$\pm$0.09   &  &  &$-$5.214 & \\
\hline
\noalign{\smallskip}
\end{tabular}
\end{flushleft}
\end{table}

\setcounter{table}{5}

\begin{table}[!hbp]
\caption[ ]{cont.} 
\font\grande=cmr7
\grande
\begin{flushleft}
\begin{tabular}{lllllrrcllllll}
\hline\noalign{\smallskip}
\multicolumn{1}{c}{$\lambda$}&
\multicolumn{1}{c}{$\log~gf$}&
\multicolumn{1}{c}{$\log~gf$}&
\multicolumn{1}{c}{source}&
\multicolumn{1}{c}{$log\gamma_{S}$}\\
\hline\noalign{\smallskip}
\multicolumn{1}{c}{stellar}&
\multicolumn{1}{c}{stellar}&
\multicolumn{1}{c}{literature}& & &\\
\hline\noalign{\smallskip}
 5044.92 & $-$0.80$\pm$0.00\\
 5080.51 & $-$0.22$\pm$0.12\\
 5122.31 &$-$0.37$\pm$0.09  &  &  &$-$4.951\\
 5188.08 &$-$1.10$\pm$0.00 \\
 5260.42 &$-$0.37$\pm$0.08 &$-$0.437 &NIST4&\\
 5261.95 &$+$0.25$\pm$0.00 &$+$0.150&NIST4&$-$5.495 &\\
 5268.25 &$-$0.80$\pm$0.12& & &$-$4.978 &\\
 5292.22 &$+$0.49$\pm$0.06 &$+$0.351 & NIST4 &$-$5.482 &\\
         &                   &$+$0.382$\pm$0.013& ZBD\\
5309.27 &$-$0.95$\pm$0.00 &\\
 5313.76 &$-$0.09$\pm$0.04 \\
 5339.355&$-$0.10$\pm$0.03& $+$0.048$\pm$0.019& ZBD \\
 5368.075&$-$1.05$\pm$0.00 \\
 5372.405&$-$0.15$\pm$0.06 & $-$0.211&NIST4&$-$5.551 \\
 5419.155&$+$0.37$\pm$0.03 & $+$0.215&NIST4  &$-$5.481 \\
         &                   & $+$0.256$\pm$0.015& ZBD\\
   5438.96 &$-$0.44$\pm$0.00  &$-$0.183& NIST4 &$-$5.544\\
    5450.45 &$-$0.97$\pm$0.09  \\
    5460.365&$-$0.77$\pm$0.04 &$-$0.673$\pm$0.030&ZBD &$-$5.531&\\
    5472.60 &$-$0.55$\pm$0.00 &$-$0.449 & NIST4&$-$5.482\\
            &                  & $-$0.362$\pm$0.030&ZBD\\
    5531.05 &$-$0.78$\pm$0.10  & $-$0.616&NIST4 &$-$5.504\\
            &                    & $-$0.632$\pm$0.021 &ZBD\\
    5616.65 &$-$0.70$\pm$0.17 &\\ 
    5659.38 &$-$0.65$\pm$0.15 & & &$-$5.407  &\\
    5667.540& $-$0.53$\pm$0.08  & &   &$-$5.535\\ 
    5699.61 & $-$0.85& \\
    5719.587& $-$0.80$\pm$0.00 &$-$0.746& NIST4&\\
            &                   &$-$0.687$\pm$0.023 & ZBD\\
    5726.88 & $-$0.28$\pm$0.05\\
    5750.99  &$-$0.40$\pm$0.05\\
    5758.665 &$-$0.35$\pm$0.00 & & & $-$5.539 \\
    5776.39 &$-$0.70 &  &  & $-$5.488\\
    5893.29 &$-$0.90 \\
    5905.115&$-$0.75$\pm$0.10 \\
    5945.53 &$-$0.67$\pm$0.09  &  & &$-$5.527\\
    5971.135 &$-$0.50 \\
    5976.460 &$-$0.29$\pm$0.06 &$-$0.222 &NIST4  &$-$5.545\\
             &                   & $-$0.317$\pm$0.023 &ZBD\\ 
    6036.170&$-$0.56$\pm$0.06  &$-$0.609 &NIST4 &$-$5.535 &  \\
            &                    &$-$0.562$\pm$0.020 &ZBD\\
    6051.120&$-$0.28$\pm$0.04  &$-$0.252& NIST4 &$-$5.515 &  \\
            &                    & $-$0.257$\pm$0.020 & ZBD\\

\hline
\noalign{\smallskip}
\end{tabular}
\end{flushleft}
\end{table}

\setcounter{table}{5}

\begin{table}[!hbp]
\caption[ ]{cont.} 
\font\grande=cmr7
\grande
\begin{flushleft}
\begin{tabular}{lllllrrcllllll}
\hline\noalign{\smallskip}
\multicolumn{1}{c}{$\lambda$}&
\multicolumn{1}{c}{$\log~gf$}&
\multicolumn{1}{c}{$\log~gf$}&
\multicolumn{1}{c}{source}&
\multicolumn{1}{c}{$log\gamma_{S}$}\\
\hline\noalign{\smallskip}
\multicolumn{1}{c}{stellar}&
\multicolumn{1}{c}{stellar}&
\multicolumn{1}{c}{literature}& & &\\
\hline\noalign{\smallskip}
    6097.57 &$-$0.39$\pm$0.06 &$-$0.237&NIST4 &\\
            &                   & $-$0.355$\pm$0.025 &ZBD\\ 
    6101.37 &$-$0.50$\pm$0.28 &\\
    6194.07 &$+$0.05$\pm$0.15 \\
    6270.81 &$-$0.18$\pm$0.12 & $-$0.196& NIST4&$-$5.510\\
    6277.54 &$---$ &            $-$0.894&NIST4& $-$5.543\\
            &      &            $-$0.778$\pm$0.021 & ZBD\\
    6300.830& $-$1.10  \\
    6343.95& $-$0.64$\pm$0.10 &$-$0.786$\pm$0.024& ZBD& \\
    6356.33& $-$0.25 \\
    6375.28 &$-$1.00 \\
    6512.79 &$-$1.00$\pm$0.00 \\
    6528.65 &$-$0.40 \\
    6594.97 & 0.00$\pm$0.00 \\
    6597.23 & $-$0.60$\pm$0.00\\
    6620.02 &$-$0.85$\pm$0.00&  \\
    6694.285&$-$0.92$\pm$0.12&$-$0.912$\pm$0.020&ZBD \\
    6788.71&$-$0.50\\
    6790.37& $-$0.70\\
    6805.74&$---$&$-$0.595&NIST4\\
           &     &$-$0.547$\pm$0.023& ZBD \\
    6990.835& $+$0.30$\pm$0.05 & $+$0.200& NIST4\\
           &                   & $+$0.084$\pm$0.032 &ZBD\\
    7082.15 & $+$0.05 \\
    7164.85&$+$0.20$\pm$0.00\\
    7284.34& $-$0.50  & \\
    7339.30 &$+$0.45?  \\
    7787.04 &$-$0.50?  \\
\hline
\noalign{\smallskip}
\end{tabular}
\end{flushleft}
\end{table}

\begin{table}[!hbp]
\caption[ ]{A few 7s, 5d, and 6d even Xe II energy levels from Hansen \& Persson (1987) modified according
to the wavelength positions observed in the UVES spectra of HR\,6000, HD\,71066,
46\,Aql, and HD\,175640} 
\begin{flushleft}
\begin{tabular}{rrlllrrcllllll}
\hline\noalign{\smallskip}
\multicolumn{2}{c}{Term}&
\multicolumn{2}{c}{level value (cm$^{-1}$)}\\
           & & NIST & This paper\\
\hline\noalign{\smallskip}
5s$^{2}$5p$^{4}$($^{3}$P$_{2}$)7s & [2]$_{5/2}$ & 132518.82 &  132519.23 \\
                                  & [2]$_{3/2}$ & 133189.42 &  133189.94 \\
5s$^{2}$5p$^{4}$($^{3}$P$_{0}$)7s & [0]$_{1/2}$ & 140883.42 &  140883.79 \\
5s$^{2}$5p$^{4}$($^{1}$D$_{2}$)5d & [0]$_{1/2}$ & 135060.97 &  135061.36 \\
5s$^{2}$5p$^{4}$($^{3}$P$_{2}$)6d & [4]$_{9/2}$ & 136109.65 &  136110.13 \\
                                  & [4]$_{7/2}$ & 136597.81 &  136598.48 \\
                                  & [3]$_{7/2}$ & 135507.32 &  135507.72 \\
                                  & [3]$_{5/2}$ & 139094.28 &  139094.83 \\
                                  & [2]$_{5/2}$ & 135547.13 &  135547.53 \\
                                  & [2]$_{3/2}$ & 135708.32 &  135708.72 \\
                                  & [1]$_{3/2}$ & 139640.43 &  139640.61 \\
                                  & [1]$_{1/2}$ & 136554.11 &  136554.47 \\
5s$^{2}$5p$^{4}$($^{3}$P$_{1}$)6d & [3]$_{7/2}$ & 145587.61 &  145588.12 \\
                                  & [3]$_{5/2}$ & 146927.86 &  146928.34 \\
                                  & [2]$_{3/2}$ & 145940.34 &  145940.79 \\
                                  & [1]$_{3/2}$ & 148085.19 &  148085.36 \\
                                  & [1]$_{1/2}$ & 145222.72 &  145223.16 \\
5s$^{2}$5p$^{4}$($^{3}$P$_{0}$)6d & [2]$_{5/2}$ & 144384.90 &  144385.45 \\
                                  & [2]$_{3/2}$ & 144140.16 &  144140.69 \\
5s$^{2}$5p$^{4}$($^{1}$D$_{2}$)6d & [4]$_{9/2}$ & 152806.73 &  152806.73 ? \\
                                  & [4]$_{7/2}$ & 152708.92 &  152709.19 \\
                                  & [1]$_{3/2}$ & 153584.09 &  153584.02 \\
\hline
\noalign{\smallskip}
\end{tabular}
\end{flushleft}
\end{table}

\begin{acknowledgements}

Kutluay Y\"uce was supported by T\"UB\.ITAK (The Scientific and Technological
Research Council of Turkey). She thanks T\"UB\.ITAK and Ankara University. 

\end{acknowledgements}

\Online

\appendix

\section{The lines used for the abundance analysis of HD\,71066}

Table\,A.1 lists the lines that were used to derive the abundances
of HD\,71066.
 The wording ``not obs'' is given for lines not present in the spectra,
while the wordings ``profile'' and ``blend'' are given for lines observed
well in the spectra, but that do not have measurable equivalent widths 
either because the noise affects the profile too much or because 
other components affect the line. These wordings also indicate
lines for which adequate equivalent widths cannot be computed, 
as in the cases of \ion{Mg}{ii} at 4481\,\AA\ which is 
a blend of transitions belonging to the same multiplet,
of most \ion{Mn}{ii} lines that are affected by hyperfine structure, of
the \ion{Ca}{ii} infrared triplet, which is a blend of isotopic components,
and so on.
For the remaining lines the measured equivelent widths are given in the table.

\begin{table*}[!hbp]
\caption[ ]{Abundances of HD\,71066 from the ATLAS12 model with parameters
\teff=12000\,K, \logg=4.1}
\font\grande=cmr7
\grande
\begin{flushleft}
\begin{tabular}{lllrrrrlcrr}
\hline\noalign{\smallskip}
 & & & & &
\multicolumn{2}{c}{HD\,71066[12000,4.1,AT12]}
\\
\hline\noalign{\smallskip}
\multicolumn{1}{c}{Species}&
\multicolumn{1}{c}{$\lambda$($\AA$)}&
\multicolumn{1}{c}{$\log\,gf$}&
\multicolumn{1}{c}{Ref.$^{a}$}&
\multicolumn{1}{c}{$\chi_{low}$}&
\multicolumn{1}{c}{W(m$\AA$)}&
\multicolumn{1}{c}{$log(N_{Z})/N_{\rm tot}$)}&
\multicolumn{1}{c}{Notes}
\\
\hline\noalign{\smallskip}
\ion{He}{i}$^{a}$  &4026.209 &$-$0.374& NIST4 &169087.008&profile& $-$2.28& the core is computed too strong\\
\ion{He}{i}  &4471.502 &$+$0.043& NIST4 &169087.008&profile& $-$2.28& the core is computed too strong\\
\ion{He}{i}  &5875.661   &$+$0.739& NIST4 &169086.964&profile& $-$2.50& the core is computed too strong\\
\ion{He}{i}  &6678.151 &$+$0.328& NIST4 &171135.00&profile& $-$2.50\\
\\
\ion{Be}{ii} & 3130.420 & $-$0.178 & NIST4 &    0.00  & profile & $-$10.80 \\  
\\
\ion{C}{ii}  & 3918.968 & $-$0.533 &NIST4&131724.370  &profile & $-$3.90 & observed at 3918.92\,\AA  \\
\ion{C}{ii}  & 4267.001 & $+$0.563 &NIST4&145549.270  &profile & $-$3.90 & observed at 4267.10\,\AA \\
\ion{C}{ii}  & 4267.261 & $+$0.716 &NIST4&145550.700  &profile & $-$3.90 & \\
\ion{C}{ii}  & 4267.261 & $-$0.584 &NIST4&145550.700  &profile & $-$3.90 & \\
\ion{C}{ii}  & 6578.052 & $-$0.021 &NIST4&116537.65  &profile & $-$3.90 & \\
\ion{C}{ii}  & 7236.420 & $+$0.294 &NIST4&131735.52  &profile & $-$3.90 & observed at 7236.35\,\AA \\
\\
\ion{N}{i}   &8680.282  & $+$0.359 &NIST4& 83364.620  & not obs& $\le$$-$5.50 \\
\ion{N}{i}   &8683.403  & $+$0.105 &NIST4& 88317.830  & not obs& $\le$$-$5.50 \\
\\
\ion{O}{i}   &4368.193  & $-$2.665 &NIST4&76794.978   &profile & $-$3.70 \\
\ion{O}{i}   &4368.242  & $-$1.964 &NIST4&76794.978   &profile & $-$3.70 \\
\ion{O}{i}   &4368.258  & $-$2.186 &NIST4&76794.978   &profile & $-$3.70 \\
\ion{O}{i}   &5329.096  & $-$1.938 &NIST4&86625.757   &profile & $-$3.67: \\
\ion{O}{i}   &5329.099  & $-$1.586 &NIST4&86625.757   &profile & $-$3.67: \\
\ion{O}{i}   &5329.107  & $-$1.695 &NIST4&86625.757   &profile & $-$3.67: \\
\ion{O}{i}   &6155.961  & $-$1.363 &NIST4&86625.757   &profile & $-$3.57 \\
\ion{O}{i}   &6155.971  & $-$1.011 &NIST4&86625.757   &profile & $-$3.57 \\
\ion{O}{i}   &6155.989  & $-$1.120 &NIST4&86625.757   &profile & $-$3.57 \\
\ion{O}{i}   &6156.737  & $-$1.487 &NIST4&86627.778   &profile & $-$3.57 \\
\ion{O}{i}   &6156.755  & $-$0.898 &NIST4&86627.778   &profile & $-$3.57 \\
\ion{O}{i}   &6156.778  & $-$0.694 &NIST4&86627.778   &profile & $-$3.57 \\
\ion{O}{i}   &6454.444  & $-$1.066 &NIST4&86627.778   &profile & $-$3.57 \\
\ion{O}{i}   &6455.977  & $-$0.920 &NIST4&86631.454  &profile  & $-$3.60 \\
\ion{O}{i}   &7002.173  & $-$2.644 &NIST4&88631.146 &profile & $-$3.58 \\
\ion{O}{i}   &7002.196  & $-$1.489 &NIST4&88631.146 &profile & $-$3.58 \\
\ion{O}{i}   &7002.230  & $-$0.741 &NIST4&88631.146 &profile & $-$3.58 \\
\ion{O}{i}   &7002.250  & $-$1.364 &NIST4&88631.303 &profile & $-$3.58 \\
\\
\ion{Ne}{i}  &6402.248  & $+$0.345 &NIST4&134041.840& not obs & $\le$$-$5.70\\
\ion{Ne}{i}  &7032.413  & $-$0.249 &NIST4&134041.840& not obs & $\le$$-$5.70\\
\\
\ion{Na}{i}  &5889.950  & $+$0.108 &NIST4& 0.00     &38.2 & $-$5.42 \\
\ion{Na}{i}  &5895.924  & $-$0.194 &NIST4& 0.00     &19.3 & $-$5.62\\
\\

\hline
\noalign{\smallskip}
\end{tabular}
\end{flushleft}
\end{table*}

\setcounter{table}{0}

\begin{table*}[!hbp]
\caption[ ]{cont.}
\font\grande=cmr7
\grande
\begin{flushleft}
\begin{tabular}{lllrrrrlcrr}
\hline\noalign{\smallskip}
 & & & & &
\multicolumn{2}{c}{HD\,71066[12000,4.1,AT12]}
\\
\hline\noalign{\smallskip}
\multicolumn{1}{c}{Species}&
\multicolumn{1}{c}{$\lambda$($\AA$)}&
\multicolumn{1}{c}{$\log\,gf$}&
\multicolumn{1}{c}{Ref.$^{a}$}&
\multicolumn{1}{c}{$\chi_{low}$}&
\multicolumn{1}{c}{W(m$\AA$)}&
\multicolumn{1}{c}{$log(N_{Z})/N_{\rm tot}$)}&
\multicolumn{1}{c}{Notes}
\\
\hline\noalign{\smallskip}

\ion{Mg}{i} &5167.321  & $-$0.870 &NIST4&21850.405 &1.30 &$-$5.36\\
\ion{Mg}{i} &5172.684  & $-$0.393 &NIST4&21870.464 &4.60 &$-$5.27\\
\\
\ion{Mg}{ii} &4481.126  & $+$0.749 &NIST4&71490.190 &profile &$-$5.40\\
\ion{Mg}{ii} &4481.150  & $-$0.553 &NIST4&71490.190 &profile &$-$5.40\\
\ion{Mg}{ii} &4481.325  & $+$0.594 &NIST4&71491.063 &profile &$-$5.40\\
\\
\ion{Al}{i}  &3944.006  & $-$0.638 &NIST4&    0.000 &not obs &$\le$$-$7.30\\
\ion{Al}{i}  &3961.520  & $-$0.336 &NIST4&  112.061 &not obs &$\le$$-$7.30 \\
\\
\ion{Al}{ii} &7056.712  & $+$0.110 &NIST4  &91274.500 &not obs &$\le$$-$7.30\\
\\
\ion{Si}{ii} &3853.665  & $-$1.341 &NIST4  &55309.350&66.7 &$-$4.81 \\
\ion{Si}{ii} &3856.018  & $-$0.406 &NIST4  &55325.180&113.7 &$-$4.91\\
\ion{Si}{ii} &3862.595  & $-$0.757 &NIST4  &55309.350&101.4 &$-$4.74\\
\ion{Si}{ii} &4072.709  & $-$2.701 &NIST4  &79338.500&2.3 &$-$4.32\\
\ion{Si}{ii} &4075.452  & $-$1.400 &NIST4  &79355.020&16.37 &$-$4.65\\
\ion{Si}{ii} &4190.724  & $-$0.351 &LA  &108820.600&8.35 &$-$4.58 \\
\ion{Si}{ii} &4198.133  & $-$0.611 &LA  &108778.700&5.98 &$-$4.48 \\
\ion{Si}{ii} &5041.024  & $+$0.029 &NIST4  &81191.340&82.89 &$-$4.35\\
\ion{Si}{ii} &5055.984  & $+$0.523 &NIST4  &81251.320&100.6 &$-$4.60 \\
\ion{Si}{ii} &5056.317  & $-$0.492 &NIST4  &81251.320&46.08 &$-$4.48\\
\ion{Si}{ii} &5957.559  & $-$0.225 &NIST4  &81191.340&44.44 &$-$4.57 \\
\ion{Si}{ii} &5978.930  & $+$0.084 &NIST4  &81251.320&56.53 &$-$4.62\\
\ion{Si}{ii} &7849.722  & $+$0.470 &NIST4  &101024.350&10.65 &$-$4.94\\
\\
\ion{P}{ii}  & 4044.576 & $+$0.481 &K,MRB & 107360.250 & 18.35 & $-$5.04\\
\ion{P}{ii}  & 4127.559 & $-$0.110 &K,KP  & 103667.860 &  8.01 & $-$5.13\\
\ion{P}{ii}  & 4288.606 & $-$0.630 &K,MRB & 101635.690 &  2.10 & $-$5.34\\
\ion{P}{ii}  & 4420.712 & $-$0.329 &NIST4 &  88893.220  & 15.97 & $-$5.12\\
\ion{P}{ii}  & 4452.472 & $-$0.194 &K,MRB & 105302.170  &  6.30 & $-$5.00\\
\ion{P}{ii}  & 4463.027 & $+$0.026 &K,MRB & 105549.670  &  8.67 & $-$5.04\\
\ion{P}{ii}  & 4466.140 & $-$0.560 &NIST4 & 105549.670  &  1.83 & $-$5.24 \\
\ion{P}{ii}  & 4475.270 & $+$0.440 &NIST4 & 105549.670  &13.15 & $-$5.20  \\
\ion{P}{ii}  & 5296.077 & $-$0.160 &NIST4 &  87124.600  &22.90 & $-$4.86\\
\ion{P}{ii}  & 5344.729 & $-$0.390 &NIST4 &  86597.550  &15.49 & $-$4.99\\
\ion{P}{ii}  & 5425.880 & $+$0.180 &NIST4 &  87124.600  &31.31 & $-$4.92\\
\ion{P}{ii}  & 6034.039 & $-$0.220 &NIST4 & 86597.550   &16.81 & $-$4.95\\
\ion{P}{ii}  & 6043.084 & $+$0.416 &NIST4 & 87124.600 &32.87 & $-$4.94\\
\\
\hline
\noalign{\smallskip}
\end{tabular}
\end{flushleft}
\end{table*}

\setcounter{table}{0}

\begin{table*}[!hbp]
\caption[ ]{cont.}
\font\grande=cmr7
\grande
\begin{flushleft}
\begin{tabular}{lllrrrrlcrr}
\hline\noalign{\smallskip}
 & & & & &
\multicolumn{2}{c}{HD\,71066[12000,4.1,AT12]}
\\
\hline\noalign{\smallskip}
\multicolumn{1}{c}{Species}&
\multicolumn{1}{c}{$\lambda$($\AA$)}&
\multicolumn{1}{c}{$\log\,gf$}&
\multicolumn{1}{c}{Ref.$^{a}$}&
\multicolumn{1}{c}{$\chi_{low}$}&
\multicolumn{1}{c}{W(m$\AA$)}&
\multicolumn{1}{c}{$log(N_{Z})/N_{\rm tot}$)}&
\multicolumn{1}{c}{Notes}
\\
\hline\noalign{\smallskip}

\ion{P}{iii} & 4222.198 &   $+$0.210 &NIST4 &117835.950     &4.99 & $-$5.13\\
\\
\ion{S}{ii}  & 4153.068 &   $+$0.617 &NIST4 &128233.200      & 2.98 & $-$5.66\\
\ion{S}{ii}  & 4162.665 &   $+$0.777 &NIST4 &128599.160      & 2.67 & $-$5.87\\
\\
\ion{Ca}{i}  & 4226.728 &   $+$0.244 &NIST4 &     0.000    &profile &$-$5.68 \\ 
\\      
\ion{Ca}{ii} & 3158.869 &   $+$0.27 &NIST4 &   25191.51   & 29.96 &$-$6.60 &\\       
\ion{Ca}{ii} & 3179.331 &   $+$0.52 &NIST4 &   25414.40   & 32.47 &$-$6.73 &\\       
\ion{Ca}{ii} & 3181.275 &   $-$0.45 &NIST4 &   25414.40   & 15.90 &$-$6.45 &\\       
\ion{Ca}{ii} & 3933.663 &   $+$0.135 &NIST4 &     0.000    & profile &$-$6.33 &\\       
\ion{Ca}{ii} & 3968.469 &   $-$0.18 &NIST4 &     0.000    & profile &$-$6.90 & \\       
\ion{Ca}{ii} & 8498.023 &   $-$1.45 &GAL &   13650.19   & profile &$-$6.33 & $\Delta\lambda$=+0.16 \\       
\ion{Ca}{ii} & 8542.091 &   $-$0.50 &GAL &   13710.88   & profile &$-$6.33 & $\Delta\lambda$=+0.16  \\       
\ion{Ca}{ii} & 8662.142 &   $-$0.76 &GAL &   13650.19   & profile &$-$6.33 & $\Delta\lambda$=+0.16 \\       
\\
\ion{Sc}{ii} & 4246.822 &   $+$0.242 &NIST4 &   2540.950    & not obs & $\le$$-$10.5  \\       
\ion{Sc}{ii} & 4314.083 &   $-$0.100 &NIST4 &   4987.790    & not obs & $\le$$-$10.5 \\       
\\
\ion{Ti}{ii} & 4163.644 &   $-$0.130 &PTP   &  20891.660 & 40.17 & $-$6.45 \\       
\ion{Ti}{ii} & 4287.873 &   $-$1.790 &PTP   &  8710.440  & 9.09 & $-$6.51 \\       
\ion{Ti}{ii} & 4290.215 &   $-$0.850 &PTP   &  9395.710  & 37.88 &$-$6.51  \\       
\ion{Ti}{ii} & 4294.094 &   $-$0.930 &PTP   &  9744.250  & 40.01 & $-$6.41 \\       
\ion{Ti}{ii} & 4300.042 &   $-$0.440 &PTP   &  9518.060  & 57.29 & $-$6.39 \\       
\ion{Ti}{ii} & 4301.922 &   $-$1.150 &PTP   &  9363.620  & 24.83 &  $-$6.55\\       
\ion{Ti}{ii} & 4367.652 &   $-$0.860 &PTP   & 20891.660  & 12.52 &  $-$6.53 \\       
\ion{Ti}{ii} & 4395.031 &   $-$0.540 &PTP   &  8744.250  &55.79 &  $-$6.38\\       
\ion{Ti}{ii} & 4399.765 &   $-$1.190 &PTP   &  9975.920  &24.64&   $-$6.94\\       
\ion{Ti}{ii} & 4411.072 &   $-$0.670 &PTP   & 24961.030  &13.25 &   $-$6.48\\       
\ion{Ti}{ii} & 4417.714 &   $-$1.190 &PTP   &  9395.710  &24.69 &   $-$6.44 \\       
\ion{Ti}{ii} & 4443.810 &   $-$0.720 &PTP   &  8710.440  &49.85 &  $-$6.37 \\       
\ion{Ti}{ii} & 4464.448 &   $-$1.810 &PTP   &  9363.620  &9.85 &   $-$6.42 \\       
\ion{Ti}{ii} & 4468.492 &   $-$0.620 &NIST4 &  9118.260  &51.86 &  $-$6.41 \\       
\ion{Ti}{ii} & 4488.325 &   $-$0.510 &PTP   & 25192.710  &16.92 &   $-$6.46 \\       
\ion{Ti}{ii} & 4805.085 &   $-$1.120 &NIST4 & 16625.110  &18.79 &   $-$6.31 \\       
\ion{Ti}{ii} & 4911.195 &   $-$0.610 &PTP   & 25192.790  &14.40 &   $-$6.44 \\ 
\\      
\ion{V}{ii}  & 3093.105 &   $+$0.559 &K10V   &  3162.800  &not obs &$\le$$-$10.0 \\       
\ion{V}{ii}  & 3102.294 &   $+$0.434 &K10V   &  2968.220  &not obs & $\le$$-$10.0\\       
\\

\hline
\noalign{\smallskip}
\end{tabular}
\end{flushleft}
\end{table*}

\setcounter{table}{0}

\begin{table*}[!hbp]
\caption[ ]{cont.}
\font\grande=cmr7
\grande
\begin{flushleft}
\begin{tabular}{lllrrrrlcrr}
\hline\noalign{\smallskip}
 & & & & &
\multicolumn{2}{c}{HD\,71066[12000,4.1,AT12]}
\\
\hline\noalign{\smallskip}
\multicolumn{1}{c}{Species}&
\multicolumn{1}{c}{$\lambda$($\AA$)}&
\multicolumn{1}{c}{$\log\,gf$}&
\multicolumn{1}{c}{Ref.$^{a}$}&
\multicolumn{1}{c}{$\chi_{low}$}&
\multicolumn{1}{c}{W(m$\AA$)}&
\multicolumn{1}{c}{$log(N_{Z})/N_{\rm tot}$)}&
\multicolumn{1}{c}{Notes}
\\
\hline\noalign{\smallskip}

\ion{Cr}{ii} & 4812.337 &   $-$1.997 &K10Cr   &  31168.580  &6.07 &$-$6.22 \\       
\ion{Cr}{ii} & 4824.127 &   $-$0.980 &K10Cr      &  31219.350  &39.36 &$-$6.06\\       
\ion{Cr}{ii} & 4836.229 &   $-$1.963 &K10Cr      &  31117.390  &7.39 &$-$6.16\\       
\ion{Cr}{ii} & 5237.329 &   $-$1.160 &NIST4   &  32854.310  &22.48 &$-$6.24 \\       
\ion{Cr}{ii} & 5246.768 &   $-$2.460 &NIST4   &  29951.880  &2.93 &$-$6.17\\       
\\
\ion{Mn}{ii} & 3917.318 &   $-$1.135 &K09Mn   &  55759.270  &profile &$-$5.93\\       
\ion{Mn}{ii} & 4363.255$^{b}$ &   $-$1.887 &K09Mn   &  44899.820  &profile &$-$5.93 \\       
\ion{Mn}{ii} & 4365.217$^{b}$ &   $-$1.344 &K09Mn   &  44899.820  &profile &$-$5.93 \\       
\ion{Mn}{ii} & 4478.637$^{b}$ &   $-$0.945 &K09Mn   &  53597.130  &profile &$-$5.93 \\       
\ion{Mn}{ii} & 4806.823 &   $-$1.571 &K09Mn     &  43696.120&profile &$-$6.03\\
\\
\ion{Fe}{i}  & 3581.193 &   $+$0.406 &FW06 & 6928.27  & 28.28 & $-$3.68 \\
\ion{Fe}{i}   & 3618.768 &   $-$0.003 &FW06 & 7985.78  &15.49 & $-$3.88 \\
\ion{Fe}{i}   & 4005.242 &   $-$0.610 &FW06 &12560.93  &14.40 & $-$3.87 \\
\ion{Fe}{i}   & 4071.738 &   $-$0.022 &FW06 &12698.55  &31.00 &  $-$3.91\\
\ion{Fe}{i}   & 4202.029 &   $-$0.708 &FW06 &11976.24  &13.44 &  $-$3.85\\
\ion{Fe}{i}   & 4219.360 &   $+$0.000 &FW06 &28819.95  & 7.80 &  $-$3.82\\
\ion{Fe}{i}   & 4235.936 &   $-$0.341 &FW06 &19562.44  &12.27 &  $-$3.81 \\
\ion{Fe}{i}   & 4271.760 &   $-$0.164 &FW06 &11976.24  &30.68 &  $-$3.84 \\
\ion{Fe}{i}   & 4383.545 &   $+$0.200 &FW06 &11976.24  &43.10 &  $-$3.85 \\
\ion{Fe}{i}   & 4404.750 &   $-$0.142 &FW06 &12560.93  &29.22 &  $-$3.87 \\
\ion{Fe}{i}   & 4415.122 &   $-$0.615 &FW06 &12968.55  &14.49 &  $-$3.84 \\
\ion{Fe}{i}   & 5364.871 &   $+$0.228 &FW06 &35856.40  &3.77  &  $-$3.96\\
\\
\ion{Fe}{ii}   & 4128.748 &   $-$3.580 &FW06 &20830.58 &31.85 &   $-$3.92 \\
\ion{Fe}{ii}   & 4178.862 &   $-$2.440 &FW06 &20830.58 &67.36 &  $-$3.96 \\
\ion{Fe}{ii}   & 4273.326 &   $-$3.300 &FW06 &21812.05 &41.90 &  $-$3.86 \\
\ion{Fe}{ii}   & 4296.572 &   $-$2.930 &FW06 &21812.05 &54.16 &  $-$3.85 \\
\ion{Fe}{ii}   & 4369.411 &   $-$3.580 &FW06 &22409.85 &27.75 &   $-$3.93 \\
\ion{Fe}{ii}   & 4413.601 &   $-$4.190 &FW06 &21581.64 &15.83&   $-$3.75 \\
\ion{Fe}{ii}   & 4416.830 &   $-$2.600 &FW06 &22409.85 &64.47 &  $-$3.83 \\
\ion{Fe}{ii}   & 4491.405 &   $-$2.640 &FW06 &23031.30 &57.57 &  $-$3.96 \\
\ion{Fe}{ii}   & 4508.288 &   $-$2.350 &FW06 &23031.30 &73.82 &  $-$3.76 \\
\ion{Fe}{ii}   & 4515.339 &   $-$2.360 &FW06 &23939.36 &65.01 &  $-$3.94 \\
\ion{Fe}{ii}   & 4913.295 &   $+$0.016 &J07  &82978.71 &33.22 &  $-$3.77 \\
\ion{Fe}{ii}   & 4993.358 &   $-$3.680 &FW06 &22637.20 &26.98 &   $-$3.83\\
\ion{Fe}{ii}   & 5001.953 &   $+$0.933 &J07  &82853.65 &65.38 &  $-$3.85 \\
\ion{Fe}{ii}   & 5030.631 &   $+$0.431 &FW06 &82978.68 &44.23  &   $-$3.87 \\
\ion{Fe}{ii}   & 5035.700 &   $+$0.630 &FW06 &82978.68 &52.34  &   $-$3.84\\

\hline
\noalign{\smallskip}
\end{tabular}
\end{flushleft}
\end{table*}

\setcounter{table}{0}

\begin{table*}[!hbp]
\caption[ ]{cont.}
\font\grande=cmr7
\grande
\begin{flushleft}
\begin{tabular}{lllrrrrlc}
\hline\noalign{\smallskip}
 & & & & &
\multicolumn{2}{c}{HD\,71066[12000,4.1,AT12]}
\\
\hline\noalign{\smallskip}
\multicolumn{1}{c}{Species}&
\multicolumn{1}{c}{$\lambda$($\AA$)}&
\multicolumn{1}{c}{$\log\,gf$}&
\multicolumn{1}{c}{Ref.$^{a}$}&
\multicolumn{1}{c}{$\chi_{low}$}&
\multicolumn{1}{c}{W(m$\AA$)}&
\multicolumn{1}{c}{$log(N_{Z})/N_{\rm tot}$)}&
\multicolumn{1}{c}{Notes}
\\
\hline\noalign{\smallskip}

\ion{Fe}{ii}      & 5144.352 &   $+$0.307 &FW06 &84424.37 &23.98 &   $-$4.24\\
\ion{Fe}{ii}      & 5247.956 &   $+$0.550 &FW06 &84938.18 &41.29 &  $-$3.88 \\
\ion{Fe}{ii}      & 5260.254 &   $+$1.090 &J07  &84863.38 &65.44 &  $-$3.84 \\
\ion{Fe}{ii}      & 5276.002 &   $-$1.900 &FW06 &25805.33 &76.52 &  $-$3.95 \\
\ion{Fe}{ii}      & 5339.592 &   $+$0.568 &J07  &84296.87 &44.50 &  $-$3.85 \\
\ion{Fe}{ii}      & 5414.852 &   $-$0.258 &J07  &84863.38 &20.80 &  $-$3.72 \\
\ion{Fe}{ii}      & 5425.257 &   $-$3.390 &FW06 &25805.33 &36.19  &  $-$3.64 \\
\ion{Fe}{ii}      & 5465.932  &  $+$0.348 &FW06 &85679.70 &38.16   &  $-$3.70\\
\ion{Fe}{ii}      &  5493.830   &  $+$0.259 &FW06 &84685.20 &33.73   &  $-$3.80 \\
\ion{Fe}{ii}      & 5506.199  &  $+$0.923 &J07 &84863.38 & 53.95     &$-$3.89       \\
\ion{Fe}{ii}      & 5510.783 & $+$0.043   &J07 &85184.77&27.35 &  $-$3.76   \\
\\
\ion{Co}{ii}      & 4160.657 & $-$1.751 & K06Co &27484.371 &blend&$\le$$-$7.88\\
\\
\ion{Ni}{ii}      & 4067.031 & $-$1.834 & K03Ni &32499.530 & blend &$\le$$-$7.90\\
\\
\ion{Cu}{ii}      & 4909.734 & $+$0.790 & K03Cu  &115568.985& not obs & $\le$$-$7.8 \\
\\

\ion{Zn}{ii}      & 4911.625 & $+$0.540 &NIST4 &96909.740  &not obs & $\le$$-$7.94\\
\\
\ion{As}{ii} & 4466.348 &  & \\
\ion{As}{ii} & 4494.230&    &\\
\ion{As}{ii} & 5105.58&    &&81508.925 & 3.74\\
\ion{As}{ii} & 5231.38&    &&79128.330 & 3.16\\
\ion{As}{ii} & 5331.23&    &&81508.925 & 7.07\\
\ion{As}{ii} & 5497.727&   &&78730.893 & 4.52& &blend\\
\ion{As}{ii} & 5558.09&    &&79128.330 & 7.11& &blend\\
\ion{As}{ii} & 5651.32&    &&81508.925& 9.29\\
\ion{As}{ii} & 6110.07&    &&82819.214& 2.32\\
\ion{As}{ii} & 6170.27&    &&79128.330& 2.62&  &blend\\
\\
\ion{Sr}{ii}      & 4077.709 & $+$0.151 &NIST4 &     0.000&32.45 & $-$8.27 \\
\\
\ion{Y}{ii}       & 3950.349 & $-$0.485 &NIST4 &840.213&17.38&$-$7.68&\\
\ion{Y}{ii}       & 4883.682 & $+$0.070 &NIST4 &8743.316&25.22&$-$7.49 \\
\ion{Y}{ii}       & 4900.120 & $-$0.090 &NIST4 &8328.041&20.57&$-$7.52 \\
\\

\hline
\noalign{\smallskip}
\end{tabular}
\end{flushleft}
\end{table*}

\setcounter{table}{0}

\begin{table*}[!hbp]
\caption[ ]{cont.}
\font\grande=cmr7
\grande
\begin{flushleft}
\begin{tabular}{lllrrrrlc}
\hline\noalign{\smallskip}
 & & & & &
\multicolumn{2}{c}{HD\,71066[12000,4.1,AT12]}
\\
\hline\noalign{\smallskip}
\multicolumn{1}{c}{Species}&
\multicolumn{1}{c}{$\lambda$($\AA$)}&
\multicolumn{1}{c}{$\log\,gf$}&
\multicolumn{1}{c}{Ref.$^{a}$}&
\multicolumn{1}{c}{$\chi_{low}$}&
\multicolumn{1}{c}{W(m$\AA$)}&
\multicolumn{1}{c}{$log(N_{Z})/N_{\rm tot}$)}&
\multicolumn{1}{c}{Notes}
\\
\hline\noalign{\smallskip}

\ion{Xe}{ii}       & 4844.33 & $+$0.49 &NIST4 &93068.440&20.72&$-$5.43 \\
\ion{Xe}{ii}       & 5292.21 & $+$0.35 &NIST4 &93068.440&19.72&$-$5.20 \\
\ion{Xe}{ii}       & 5419.14 & $+$0.21 &NIST4 &95064.38&14.27&$-$5.24 \\
\ion{Xe}{ii}       & 5438.97 & $-$0.19 &NIST4 &102799.07&2.93&$-$5.55 \\
\ion{Xe}{ii}       & 5472.61 & $-$0.45 &NIST4 &95437.67&5.03&$-$5.34 \\
\ion{Xe}{ii}       & 5531.06 & $-$0.62 &NIST4 &95437.67&1.87&$-$5.71 \\
\ion{Xe}{ii}       & 5719.61 & $-$0.74 &NIST4 &96033.48&1.40&$-$5.64 \\
\ion{Xe}{ii}       & 5976.46 & $-$0.22 &NIST4 &95064.38&4.70&$-$5.49 \\
\ion{Xe}{ii}       & 6036.20 & $-$0.61 &NIST4 &95396.74&2.44&$-$5.45 \\
\ion{Xe}{ii}       & 6051.15 & $-$0.25 &NIST4 &95437.67&4.59&$-$5.44 \\
\ion{Xe}{ii}       & 6097.59 & $-$0.24 &NIST4 &95436.74&3.93&$-$5.53 \\
\ion{Xe}{ii}       & 6990.88 & $+$0.20 &NIST4 &99409.99&5.18&$-$5.36 \\
\\
\ion{Nd}{iii}      & 4927.488& $-$0.83&DREAM & 3715. & 1.78 & $-$9.63\\
\ion{Nd}{iii}      & 5294.113& $-$0.65&DREAM &    0. & 4.12 & $-$9.62\\
\\
\ion{Dy}{iii}      & 3930.640 &$-$0.88&DREAM&     0. & profile &$-$9.90\\
\\
\ion{Au}{ii}     & 4016.067 &$-$1.88 &RW& 84510.894& 2.39&$-$7.15\\
\ion{Au}{ii}     & 4052.790 &$-$1.69 &RW& 84510.894& 3.99&$-$7.08\\
\\
\ion{Hg}{i}      & 4358.314 & $-$0.321 &NIST4 &39412.300&profile&$-$6.40\\
\\
\ion{Hg}{ii}      & 3983.890 & $-$1.51 &NIST4 &35514.000&profile&$-$6.40 \\
\ion{Hg}{ii}      & 5677.102 & $+$0.82 &NIST4&105543.000& 5.56 &$-$6.19&blend \\
\hline
\noalign{\smallskip}
\end{tabular}
\end{flushleft}

$^{a}$ \ion{He}{i} profiles were compute as described in Castelli \& Hubrig (2004a).
The wavelengths and $\log\,gf$-values are multiplet values.\\
$^{b}$ The hyperfine structure was considered in the line profile computations.\\
DREAM: Bi\'emont et al.(1999): http://w3.umons.ac.be/~astro/dream.shtml;\\
NIST4: NIST Atomic Spectra Database, version 4 at http://physics.nist.gov/pml/data/asd.cfm;\\
FW06: Fuhr \& Wiese (2006); GAL: Gallagher (1967); LA: Lanz \& Artru (1985); PTP: Pickering et al. (2002);\\ 
J07: Johansson (2007);\\
K03Ni: http://kurucz.harvard.edu/atoms/2801/gf2801.pos;\\
K03Cu: http://kurucz.harvard.edu/atoms/2901/gf2901.pos;\\
K06Co: http://kurucz.harvard.edu/atoms/2701/gf2701.pos;\\
K09Mn: http://kurucz.harvard.edu/atoms/2501/gf2501.pos;\\
K10V: http://kurucz.harvard.edu/atoms/2301/gf2301.pos;\\
K10Cr: http://kurucz.harvard.edu/atoms/2401/gf2401.pos;\\
``K'' before another $\log\,gf$ source means that the $\log\,gf$ is from the Kurucz files
available at http://kurucz.harvard.edu/linelists/gf100/; in particular: 
KP: Kurucz \& Peytremann (1975); MRB: Miller et al. (1971);\\
RW: Rosberg \& Wyart (1997).\\

\end{table*}

\section{The investigated \ion{Xe}{ii} lines in HR\,6000, HD\,71066,
46\,Aql, and HD\,175640}

Table\,B.1 gives the details on the determination of the \ion{Xe}{ii} wavelengths
and $\log\,gf$-values from the spectra of the four stars. It
lists in successive columns the laboratory wavelengths and the line intensity 
taken from the NIST database (footnote\,6), the stellar wavelengths as derived from HR\,6000, HD\,71066,
46\,Aql, and HD\,175640.  
If the observed wavelength is the same in all the stars, only
that of HR\,6000 is given.
HR\,6000, HD\,71066, 46\,Aql, and HD\,175640 are indicated  in col.\,6 with the 
numbers 1,2, 3, and 4, respectively.
 A question mark means uncertain
determinations from that star.  
The wavelength difference $\Delta\lambda$=$\lambda$(stellar)-$\lambda$(lab) is given in col.\,4.
 The energy and the  configuration of the lower and
 upper level of the transition are given in cols. 7,8,9, and 10, respectively.
 The last column
adds some notes about the observed lines. Table\,B.1 lists also
the NIST 
$\log\,gf$-values and  the $\log\,gf$-values derived from the 
experimental transition rates determined
by Z\'ieli\'nska et al. (2002).

\begin{table*}[!hbp]
\caption[ ]{Xe II lines examined in HR 6000, HD\,71066, 46\,Aql, and HD\,175640.
The stars are indicated in col.\,6 with the numbers 1,2,3, and 4, respectively.
The ``N'' in column\,5 indicates that the $\log\,gf$-value 
was taken from the NIST database, while ``ZBD'' indicates data from Z\'ieli\'nska et al. (2002)} 
\font\grande=cmr7
\grande
\begin{flushleft}
\begin{tabular}{rrlllrrcllllll}
\hline\noalign{\smallskip}
\multicolumn{1}{c}{$\lambda$(Lab)}&
\multicolumn{1}{c}{Int.}&
\multicolumn{1}{c}{$\lambda$(stellar)}&
\multicolumn{1}{c}{$\Delta\lambda$}&
\multicolumn{1}{c}{$\log~gf$}&
\multicolumn{1}{c}{}&
\multicolumn{1}{c}{$\chi_{low}$(cm$^{-1}$)}&
\multicolumn{2}{c}{Term}&
\multicolumn{1}{c}{$\chi_{up}$(cm$^{-1}$)}&
\multicolumn{2}{c}{Term}&
\multicolumn{1}{l}{Notes}\\
\hline\noalign{\smallskip}
    3907.91& 100&  3907.820 & $-$0.09&$-$0.75 &1& 113512.36 &($^{3}$P$_{2}$)6p & [3]$_{5/2}$&139094.28& ($^{3}$P$_{2}$)6d&[3]$_{5/2}$&\\
           &    &           & &$-$0.80 &2&           &                  &                      &                  &           &      &blend\\
           &    &           & &$-$0.90 &3&           &                  &                      &                  &           &      &blend\\
    4037.29& 100&  4037.260 &$-$0.03& $-$1.00 &1,2,3& 111792.17 &($^{3}$P$_{2}$)6p & [2]$_{3/2}$&136554.11& ($^{3}$P$_{2}$)6d&[1]$_{1/2}$  &broad weak blend\\
    4037.59& 200&  4037.470 &$-$0.12& $-$0.75 &1,2,3& 121179.80 &($^{3}$P$_{1}$)6p & [0]$_{1/2}$&145940.34& ($^{3}$P$_{1}$)6d&[2]$_{3/2}$  &broad weak blend\\
    4057.46& 200&  4057.360  &$-$0.10& $-$0.80:&1,2?,3 & 111958.89 &($^{3}$P$_{2}$)6p & [2]$_{5/2}$&136597.81& ($^{3}$P$_{2}$)6d&[4]$_{7/2}$& blend\\
    4158.04& 200&  4157.980 &$-$0.06& $-$0.60 &1,2?,3   & 121179.80 &($^{3}$P$_{1}$)6p & [0]$_{1/2}$&145222.72& ($^{3}$P$_{1}$)6d&[1]$_{1/2}$&  blend\\
    4162.16&  60&  4162.160 &$+$0.00& $-$1.60 &1 & 107904.50 &($^{3}$P$_{1}$)5d & [1]$_{3/2}$&131923.79& ($^{1}$D$_{2}$)6p&[2]$_{3/2}$ &blend,weak,3 noise\\
           &    &           && $-$1.55 &2&\\
    4180.10&1000&  4180.007 &$-$0.093& $-$0.35N &1,2,3 & 129667.35 &($^{1}$D$_{2}$)6p & [1]$_{3/2}$&153584.09& ($^{1}$D$_{2}$)6d&[1]$_{3/2}$& blend\\
    4193.15& 500&  4193.100 & $-$0.05 &$-$0.60&1 & 128867.20 &($^{1}$D$_{2}$)6p & [3]$_{5/2}$&152708.92& ($^{1}$D$_{2}$)6d&[4]$_{7/2}$& \\
    4208.48& 400&  4208.391 &$-$0.089& $-$0.40&1,4 & 111792.17 &($^{3}$P$_{2}$)6p & [2]$_{3/2}$&135547.13& ($^{3}$P$_{2}$)6d&[2]$_{5/2}$& \\
           &    &           && $-$0.36&2,3\\
    4209.47& 200&  4209.370 &$-$0.10& $-$0.70&1,2,3,4?& 111958.89 &($^{3}$P$_{2}$)6p & [2]$_{5/2}$&135708.32& ($^{3}$P$_{2}$)6d&[2]$_{3/2}$&   4 blend\\
    4213.72& 400&  4213.620 &$-$0.10& $-$0.30&1 & 120414.87 &($^{3}$P$_{0}$)6p & [1]$_{1/2}$&144140.16& ($^{3}$P$_{0}$)6d&[2]$_{3/2}$ &blend\\
           &    &           & &$-$0.25&2,4&\\ 
          &    &           & &$-$0.08& 3&\\ 
    4215.60& 200&  4215.620 &$+$0.02& $-$1.05&1,2,3 &  93068.44 &($^{3}$P$_{2}$)6s & [2]$_{5/2}$&116783.09& ($^{3}$P$_{2}$)6p&[1]$_{3/2}$&   blend\\
    4223.00& 400&  4222.900&$-$0.10& $+$0.55&1& 123254.60 &($^{3}$P$_{1}$)6p & [2]$_{3/2}$&146927.86& ($^{3}$P$_{1}$)6d&[3]$_{5/2}$& \\
           &    &          & &$+$0.85&2,3&           &                  &           &          &                  &           &         &\\
           &    &          & &$+$0.30 &4\\
    4238.25& 500&  4238.135 &$-$0.115& $-$0.18&1& 111958.89 &($^{3}$P$_{2}$)6p & [2]$_{5/2}$&135547.13& ($^{3}$P$_{2}$)6d&[2]$_{5/2}$\\
           &    &           && $-$0.13&2&\\ 
           &    &           && $-$0.23&3\\ 
           &    &           && $-$0.40&4\\ 
    4245.38& 500&  4245.300 &$-$0.08& $-$0.08&1 &  111958.89 &($^{3}$P$_{2}$)6p & [2]$_{5/2}$&135507.32& ($^{3}$P$_{2}$)6d&[3]$_{7/2}$\\
           &    &           && $-$0.10&2,3&             &                  &                      &                              &         &   \\     
           &    &           && $-$0.25&4&             &                  &                      &                              &         &   \\     
    4251.57& 100&  4251.540&$-$0.03& $-$0.60&1? & 124571.09 &($^{3}$P$_{1}$)6p & [1]$_{1/2}$&148085.19& ($^{3}$P$_{1}$)6d&[1]$_{3/2}$& blend\\
           &    &          && $-$0.55&2? &           &                  &            &         &                  &           &          blend \\       
    4296.40& 500&  4296.320&$-$0.08 & $-$0.85&1,2,3,4 & 111792.17 &($^{3}$P$_{2}$)6p & [2]$_{3/2}$&135060.97& ($^{1}$D$_{2}$)5d&[0]$_{1/2}$\\
    4330.52&1000&  4330.390 &$-$0.13& $+$0.30 &1,2,3,4& 113512.36 &($^{3}$P$_{2}$)6p & [3]$_{5/2}$&136597.81& ($^{3}$P$_{2}$)6d&[4]$_{7/2}$\\
           &    &           & & $+$0.498N&\\
    4369.20& 200&  4369.100 &$-$0.10& $-$0.75& 1 & 113672.89 &($^{3}$P$_{2}$)6p & [1]$_{1/2}$&136554.11& ($^{3}$P$_{2}$)6d&[1]$_{1/2}$ \\
           &    &           & &$-$0.70& 2?,3  &           &                  &            &         &                  &           &         2 blend \\
    4373.78& 100&  4373.700 &$-$0.08& $-$0.70 &1,2?,3?& 116783.09 &($^{3}$P$_{2}$)6p & [1]$_{3/2}$&139640.43& ($^{3}$P$_{2}$)6d&[1]$_{3/2}$ &blend\\
    4384.93&  60&  4384.91  &$-$0.02& $-$1.95  & 1,3  &  90873.83 &5s5p$^{6}$ &$^{2}$S$_{1/2}$&113672.89& ($^{3}$P$_{2}$)6p&[1]$_{1/2}$ & blend\\
           &    &        &   &$\le$ $-$2.50&  2      &           &           &               &         &                  &   & not observed  \\
    4393.20& 500&  4393.090&$-$0.11 & $+$0.00  &1,2,3,4? & 121628.82 &($^{3}$P$_{0}$)6p & [1]$_{3/2}$&144384.90& ($^{3}$P$_{0}$)6d&[2]$_{5/2}$&\\
    4395.77& 500&  4395.770 : &~0.00&$+$0.00&1?,2?,3?& 130063.96 &($^{1}$D$_{2}$)6p & [3]$_{7/2}$&152806.73& ($^{1}$D$_{2}$)6d&[4]$_{9/2}$& blend\\

\hline
\noalign{\smallskip}
\end{tabular}
\end{flushleft}
\end{table*}

\setcounter{table}{0}

\begin{table*}[!hbp]
\caption[ ]{cont.} 
\font\grande=cmr7
\grande
\begin{flushleft}
\begin{tabular}{rrlllrrcllllll}
\hline\noalign{\smallskip}
\multicolumn{1}{c}{$\lambda$(Lab)}&
\multicolumn{1}{c}{Int.}&
\multicolumn{1}{c}{$\lambda$(stellar)}&
\multicolumn{1}{c}{$\Delta\lambda$}&
\multicolumn{1}{c}{$\log~gf$}&
\multicolumn{1}{c}{}&
\multicolumn{1}{c}{$\chi_{low}$(cm$^{-1}$)}&
\multicolumn{2}{c}{Term}&
\multicolumn{1}{c}{$\chi_{up}$(cm$^{-1}$)}&
\multicolumn{2}{c}{Term}&
\multicolumn{1}{l}{Notes}\\
\hline\noalign{\smallskip}

    4414.84& 300&  4414.84 &~0.00& $-$0.50   & 1,2,3& 109563.14 &($^{1}$D$_{2}$)6p & [3]$_{7/2}$&132207.76& ($^{1}$D$_{2}$)6p&[2]$_{5/2}$ &2,4 blend \\
           &    &          & & $+$0.243N  &&\\ 
    4416.07&150&4416.090  &$+$0.02& $-$0.80  & 1 ?& 124289.45   &($^{3}$P$_{1}$)6p & [1]$_{3/2}$ & 146927.86 & ($^{3}$P$_{1}$)6d & [3]$_{5/2}$&  3 noise\\
    4448.13& 500&  4448.025&$-$0.105& $+$0.05  &1,4 & 123112.54 &($^{3}$P$_{1}$)6p & [2]$_{5/2}$&145587.61& ($^{3}$P$_{1}$)6d&[3]$_{7/2}$&\\
           &    &          && $+$0.15 &2,3&\\
    4462.19&1000&  4462.090&$-$0.10& $+$0.33  &1,2,3 & 113705.40 &($^{3}$P$_{2}$)6p & [3]$_{7/2}$&136109.65& ($^{3}$P$_{2}$)6d&[4]$_{9/2}$& 4 blend\\
    $----$\\
    4787.77 & 100 & 4787.77 &~0.00&$-$0.88    & 1 &  111326.96 &($^{3}$P$_{1}$)5d & [2]$_{3/2}$&132207.76& ($^{1}$D$_{2}$)6p&[2]$_{5/2}$ & noise?\\
            &     &      &   &$-$0.80    &2,3&\\ 
    4818.02 & 200 & 4817.98 &$-$0.04&$-$1.25    & 1,2,4 & 96033.48 &($^{3}$P$_{2}$)5d & [2]$_{3/2}$&116783.09& ($^{3}$P$_{2}$)6p&[1]$_{3/2}$& 3 artifact \\
    4823.35 & 300 & 4823.25 &$-$0.10&$-$0.65    & 1,2,3 & 111792.17 &($^{3}$P$_{2}$)6p & [2]$_{3/2}$&132518.82& ($^{3}$P$_{2}$)7s&[2]$_{5/2}$ & 4  blend\\
    4844.33 &2000 & 4844.33 &~0.00&$+$ 0.65   & 1 &  93068.44 &($^{3}$P$_{2}$)6s & [2]$_{5/2}$&113705.40& ($^{3}$P$_{2}$)6p&[3]$_{7/2}$&\\
            &     &         & &$+$0.60&2,3,4&\\
            &     &         & &$+$0.491N&\\
            &     &         & &\multicolumn{3}{l}{$+$0.510$\pm$0.027ZBD}\\
    4876.50 & 500 & 4876.50 &~0.00& $+$0.10  & 1,2,3& 109563.14 &($^{1}$D$_{2}$)6s & [2]$_{5/2}$&130063.96& ($^{1}$D$_{2}$)6p&[3]$_{7/2}$& 4 blend \\
            &     &       &  &$+$0.255N&\\
    4883.53 & 600 & 4883.53&~0.00 &$-$0.25&1,2,3& 101157.48 &($^{3}$P$_{0}$)6s & [0]$_{1/2}$&121628.82& ($^{3}$P$_{0}$)6p&[1]$_{3/2}$ & \\    
    4884.15 & 100 & 4884.09 &$-$0.06&$-$0.80 &1   & 120414.87 &($^{3}$P$_{0}$)6p & [1]$_{1/2}$&140883.42& ($^{3}$P$_{0}$)7s&[0]$_{1/2}$ & 2,3,4 not obs. \\    
    4887.30 & 300 & 4887.30 &~0.00&$-$0.90 &1,4 & 102799.07 &($^{3}$P$_{1}$)6s & [1]$_{3/2}$&123254.60& ($^{3}$P$_{1}$)6p&[2]$_{3/2}$ &\\
            &     &        & &$-$0.80&2,3&\\
    4890.090 & 300 & 4890.085&$-$0.005 &$-$1.20  &1,3,4&   93068.44 &($^{3}$P$_{2}$)6s & [2]$_{5/2}$&113513.36& ($^{3}$P$_{2}$)6p&[3]$_{5/2}$& \\
            &     &       &   &$-$1.10&2&\\
            &     &       &   &\multicolumn{3}{l}{$-$0.754$\pm$0.022ZBD}\\
    4919.66 & 200 & 4919.66 &~0.00&   $-$0.95 &1,2& 04250.06 &($^{3}$P$_{1}$)5d & [1]$_{1/2}$&124571.09& ($^{3}$P$_{1}$)6p&[1]$_{1/2}$ \\ 
            &     &     &     & $-$0.65 &3&\\ 
            &     &     &     & $-$0.85 &4&\\ 
  4921.48 & 800 &4921.48&~0.00&   $+$ 0.10&1,2,3 & 102799.07 &($^{3}$P$_{1}$)6s & [1]$_{3/2}$&123112.54& ($^{3}$P$_{1}$)6p&[2]$_{5/2}$ &  \\
            &    &       &   &$-$0.10& 4&\\
    4971.71& 200 & 4971.68&$-$0.03& $-$0.75   &1,2,3,4 &  119085.49. &($^{1}$D$_{2}$)5d & [3]$_{5/2}$&139193.80& ($^{3}$P$_{2}$)7p&[1]$_{3/2}$&  2,3,4 noise \\
    4972.71& 400 & 4972.70&$-$0.01& $-$0.55   &1,2,3,4 &  109563.14 &($^{1}$D$_{2}$)6s & [2]$_{5/2}$&129667.35& ($^{1}$D$_{2}$)6p&[1]$_{3/2}$ \\
    4988.77& 300 & 4988.725 &$-$0.045&$-$1.00  &1  &  104250.06 &($^{3}$P$_{1}$)5d & [1]$_{1/2}$&124289.45& ($^{3}$P$_{1}$)6p&[1]$_{3/2}$& blend\\
           &     &        &  &$-$0.80&2,3,4&\\  
    5044.92& 150 & 5044.92 &~0.00&   $-$0.80  &1,2,4&  112924.84 &($^{1}$D$_{2}$)6s & [2]$_{3/2}$&132741.15& ($^{1}$D$_{2}$)6p&[1]$_{1/2}$  & 3,4 noise\\
    5080.62& 600 & 5080.51&$-$0.11&  $-$0.30  &1,3 &  113512.36 &($^{3}$P$_{2}$)6p & [3]$_{5/2}$&133189.42& ($^{3}$P$_{2}$)7s&[2]$_{3/2}$ &\\
           &     &        &    &$-$0.05 & 2&\\
    5122.42& 200 & 5122.31 &$-$0.11  &$-$0.50  & 1 & 113672.89 &($^{3}$P$_{2}$)6p & [1]$_{1/2}$&133189.42& ($^{3}$P$_{2}$)7s&[2]$_{3/2}$  & 4 noise\\
           &     &      &     &$-$0.30& 2,3 &           &                  &            &         &                  &                   & 3 blend \\
    5188.04 &200 & 5188.08 &$+$0.04  &$-$1.10 & 1,3&   123112.54  &($^{3}$P$_{1}$)6p & [2]$_{5/2}$&142382.13& ($^{3}$P$_{1}$)7s&[1]$_{3/2}$&   2 blend, 4 not obs\\

\hline
\noalign{\smallskip}
\end{tabular}
\end{flushleft}
\end{table*}

\setcounter{table}{0}

\begin{table*}[!hbp]
\caption[ ]{cont.} 
\font\grande=cmr7
\grande
\begin{flushleft}
\begin{tabular}{rrlllrrcllllll}
\hline\noalign{\smallskip}
\multicolumn{1}{c}{$\lambda$(Lab)}&
\multicolumn{1}{c}{Int.}&
\multicolumn{1}{c}{$\lambda$(stellar)}&
\multicolumn{1}{c}{$\Delta\lambda$}&
\multicolumn{1}{c}{$\log~gf$}&
\multicolumn{1}{c}{}&
\multicolumn{1}{c}{$\chi_{low}$(cm$^{-1}$)}&
\multicolumn{2}{c}{Term}&
\multicolumn{1}{c}{$\chi_{up}$(cm$^{-1}$)}&
\multicolumn{2}{c}{Term}&
\multicolumn{1}{l}{Notes}\\
\hline\noalign{\smallskip}

    5260.44 & 200  & 5260.42&$-$0.02   &$-$0.437N &1,4 &  104250.06 &($^{3}$P$_{1}$)5d & [1]$_{1/2}$&123254.60& ($^{3}$P$_{1}$)6p&[2]$_{3/2}$ &\\
            &      &    &       &$-$0.25 &2&\\
            &      &5260.44&~0.00    &$-$0.35 &3&\\ 
    5261.95&200  & 5261.95&~0.00   &$+$0.25 & 1,2,3&  112924.84 &($^{1}$D$_{2}$)6s & [2]$_{3/2}$&131923.79& ($^{1}$D$_{2}$)6p&[2]$_{3/2}$&\\
           &     &       &    &$+$0.150N&\\ 
    5268.31&50   & 5268.25&$-$0.06   &$-$1.00  &1&  105313.33 &($^{3}$P$_{2}$)5d & [1]$_{3/2}$&124289.45& ($^{3}$P$_{1}$)6p&[1]$_{3/2}$ &\\
           &     &       &    &$-$0.80 &2&\\
           &     &       &    &$-$0.70 &3,4&\\
    5292.22&1000 & 5292.22&~0.00   &$+$0.60   &1 &  93068.44 &($^{3}$P$_{2}$)6s & [2]$_{5/2}$&111958.89& ($^{3}$P$_{2}$)6p&[2]$_{5/2}$ &\\
           &     &        &   &$+$0.45& 2,3,4&\\
           &     &        &   &$+$0.351N&\\
           &     &        &   &\multicolumn{3}{l}{$+$0.382$\pm$0.013ZBD}\\
    5309.27& 200 & 5309.27&~0.00   &$-$0.95 &1,2,3,4 &  102799.07 &($^{3}$P$_{1}$)6s & [1]$_{3/2}$&121628.82& ($^{3}$P$_{0}$)6p&[1]$_{3/2}$ &\\
    5313.87& 800 & 5313.76 &$-$0.11  &$-$0.10  &1 &  113705.40 &($^{3}$P$_{2}$)6p & [3]$_{7/2}$&132518.82& ($^{3}$P$_{2}$)7s&[2]$_{5/2}$ & \\
           &     &         &  &$-$0.15  &2& \\
           &     &         &  &$-$0.05  &3,4& \\
    5339.33& 1000& 5339.355&$+$0.025  &$-$0.07   &1,4&   93068.94 &($^{3}$P$_{2}$)6s & [2]$_{5/2}$&111792.17& ($^{3}$P$_{2}$)6p&[2]$_{3/2}$ & \\
           &     &        &   &$-$0.10   &2&\\
           &     &        &   &$-$0.15   &3&\\
           &     &        &   &\multicolumn{3}{l}{$+$0.048$\pm$0.019ZBD}\\
    5368.07& 100 & 5368.075&$+$0.005&  $-$1.05   &1,2,3&  105947.55 &($^{3}$P$_{2}$)5d & [0]$_{1/2}$&124571.09& ($^{3}$P$_{1}$)6p&[1]$_{1/2}$ & &   \\
    5372.39& 300 & 5372.405&$+$0.015  &$-$0.211N &1,4 & 95064.38 &($^{3}$P$_{2}$)6s & [2]$_{3/2}$ & 113672.89 & ($^{3}$P$_{2}$)6p&[1]$_{1/2}$ &  blend\\
           &     &        &   &$-$0.10   &2,3&\\
    5419.15&2000 & 5419.155&$+$0.005  &$+$0.42 &1&  95064.38 &($^{3}$P$_{2}$)6s & [2]$_{3/2}$&113512.36& ($^{3}$P$_{2}$)6p&[3]$_{5/2}$ & \\
           &     &      &     &$+$0.35 &2,3,4&\\
           &     &      &     &$+$0.215N&\\
           &     &      &     &\multicolumn{3}{l}{$+$0.256$\pm$0.015ZBD}\\
    5438.96& 400 & 5438.96   &~0.00&$-$0.44  &1,2,3,4& 102799.07 &($^{3}$P$_{1}$)6s & [1]$_{3/2}$&121179.80& ($^{3}$P$_{1}$)6p&[0]$_{1/2}$ &\\
           &     &         &  &$-$0.183N\\
    5450.45& 100 & 5450.45&~0.00   &$-$1.10   &1&  105947.55 &($^{3}$P$_{2}$)5d & [0]$_{1/2}$&124289.45& ($^{3}$P$_{1}$)6p&[1]$_{3/2}$ &  & \\
           &     &        &   &$-$0.90   &2,3&\\
    5460.39& 300 & 5460.365&$-$0.025  &$-$0.85   &1&   95396.74 &($^{3}$P$_{2}$)5d & [2]$_{5/2}$&113705.40& ($^{3}$P$_{2}$)6p&[3]$_{7/2}$ &\\
           &     &         &  &$-$0.75   &2,3,4&\\
           &     &         &  &\multicolumn{3}{l}{$-$0.673$\pm$0.030ZBD}\\  
    5472.61& 500 & 5472.60 &$-$0.01  &$-$0.55 &1,2,3,4&  95437.67 &($^{3}$P$_{2}$)5d & [3]$_{7/2}$&113705.40& ($^{3}$P$_{2}$)6p&[3]$_{7/2}$ &\\
           &     &         &  &$-$0.449N&\\ 
           &     &         &  &\multicolumn{3}{l}{$-$0.362$\pm$0.030ZBD}\\

\hline
\noalign{\smallskip}
\end{tabular}
\end{flushleft}
\end{table*}

\setcounter{table}{0}

\begin{table*}[!hbp]
\caption[ ]{cont.} 
\font\grande=cmr7
\grande
\begin{flushleft}
\begin{tabular}{rrlllrrcllllll}
\hline\noalign{\smallskip}
\multicolumn{1}{c}{$\lambda$(Lab)}&
\multicolumn{1}{c}{Int.}&
\multicolumn{1}{c}{$\lambda$(stellar)}&
\multicolumn{1}{c}{$\Delta\lambda$}&
\multicolumn{1}{c}{$\log~gf$}&
\multicolumn{1}{c}{}&
\multicolumn{1}{c}{$\chi_{low}$(cm$^{-1}$)}&
\multicolumn{2}{c}{Term}&
\multicolumn{1}{c}{$\chi_{up}$(cm$^{-1}$)}&
\multicolumn{2}{c}{Term}&
\multicolumn{1}{l}{Notes}\\
\hline\noalign{\smallskip}

    5531.07& 400 & 5531.05 &$-$0.02 &$-$0.90  &1&   95437.67 &($^{3}$P$_{2}$)5d & [3]$_{7/2}$&113512.36& ($^{3}$P$_{2}$)6p&[3]$_{5/2}$ &\\
           &     &       &  &$-$0.80  &2,3&\\
           &     &       &  &$-$0.616N& 4\\
           &     &       &  &\multicolumn{3}{l}{$-$0.632$\pm$0.021ZBD}\\
    5616.67&150  & 5616.65 &$-$0.02 &$-$0.80 &1,3,4& 105313.33 &($^{3}$P$_{2}$)5d & [1]$_{3/2}$&123112.54& ($^{3}$P$_{1}$)6p&[2]$_{5/2}$&   blend\\ 
           &     &       &  & $-$0.40&2&\\
    5659.38 &150  & 5659.38&~0.00 &$-$0.80 &1,3& 106906.12 &($^{3}$P$_{1}$)6s & [1]$_{1/2}$&124571.09& ($^{3}$P$_{1}$)6p&[1]$_{1/2}$  &\\
           &     &        &  &$-$0.50&2,4&            &                  &            &         &                  &                 & noise ? \\
    5667.56 &300  & 5667.540&$-$0.02& $-$0.65  &1 &   96033.48 &($^{3}$P$_{2}$)5d & [2]$_{3/2}$&113672.89& ($^{3}$P$_{2}$)6p&[1]$_{1/2}$\\ 
            &     &         & &$-$0.50   &2,3 \\
            &     &         & &$-$0.45   &4 \\
    5699.61 &100  & 5699.61 &~0.00& $-$0.85?&1 & 111326.96 &($^{3}$P$_{1}$)5d & [2]$_{3/2}$&128867.20& ($^{1}$D$_{2}$)6p&[3]$_{5/2}$&  2,3,4 noise \\
    5719.61 &200  & 5719.587&$-$0.023& $-$0.80  &1,2&   96033.48 &($^{3}$P$_{2}$)5d & [2]$_{3/2}$&113512.36& ($^{3}$P$_{2}$)6p&[3]$_{5/2}$ & 3,4 blend telluric\\
           &     &       &   &$-$0.746N&\\
           &     &       &   &\multicolumn{3}{l}{$-$0.687$\pm$0.023ZBD}\\
    5726.91 &200  & 5726.88&$-$0.03 & $-$0.35  &1&  114751.08 &($^{3}$P$_{2}$)5d & [3]$_{5/2}$ &132207.76& ($^{1}$D$_{2}$)6p& [2]$_{5/2}$ &   3 blend telluric\\
           &     &      &   &$-$0.25  &2,4&\\
    5751.03 &200  & 5750.99&$-$0.04 &$-$0.35 &1,2&  106906.12 &($^{3}$P$_{1}$)6s & [1]$_{1/2}$&124289.45& ($^{3}$P$_{1}$)6p&[1]$_{3/2}$&  2 noise\\
           &      &       &  &$-$0.45& 3,4&\\
    5758.65 &100  & 5758.665&$+$0.015 &$-$0.35  &1,4&  112703.64 &($^{3}$P$_{1}$)5d & [2]$_{5/2}$&130063.96& ($^{1}$D$_{2}$)6p&[3]$_{7/2}$ & blend,  2,3 noise\\
    5776.39 &100  & 5776.39&~0.00 &$-$0.70  &1&  105947.55 &($^{3}$P$_{2}$)5d & [0]$_{1/2}$&123254.60& ($^{3}$P$_{1}$)6p&[2]$_{3/2}$ &2,3 not obs, 4 no spectrum\\
    5893.29 &150  & 5893.29 &~0.00&$-$0.90  &1&  112703.64 &($^{3}$P$_{1}$)5d & [2]$_{5/2}$&129667.35& ($^{1}$D$_{2}$)6p&[1]$_{3/2}$ &  2 noise, 3 not obs, 4 blend\\
    5905.13 &200  & 5905.115&$-$0.015 &$-$0.85  &1&  104250.06 &($^{3}$P$_{1}$)5d & [1]$_{1/2}$&121179.80& ($^{3}$P$_{1}$)6p&[0]$_{1/2}$ & &\\
           &     &         &  &$-$0.65  &2&            &                  &            &         &                  &            &  2,3 blend telluric\\
    5945.53 &300  & 5945.53&~0.00  &$-$0.60  &2,4 &   96958.18 &($^{3}$P$_{2}$)5d & [1]$_{1/2}$&113672.89& ($^{3}$P$_{2}$)6p&[1]$_{1/2}$ & 1 blend telluric\\
            &     &        &  &$-$0.80  &3 &\\ 
    5971.13 &200  & 5971.135&$+$0.005 &$-$0.50  &1 &  112924.84 &($^{1}$D$_{2}$)6s & [2]$_{3/2}$&129667.35& ($^{1}$D$_{2}$)6p&[1]$_{3/2}$  & 2,3,4 noise\\
    5976.46 &1000 & 5976.460 &$+$0.00 &$-$0.222N  &1,2&   95064.38 &($^{3}$P$_{2}$)6s & [2]$_{3/2}$&111792.17& ($^{3}$P$_{2}$)6p&[2]$_{3/2}$ &\\
            &     &      &    &$-$0.35    &3,4&\\
            &     &      &    &\multicolumn{3}{l}{$-$0.317$\pm$0.023ZBD}\\ 
    6036.20 &500  & 6036.170&$-$0.03 &$-$0.57  &1&   95396.74 &($^{3}$P$_{2}$)5d & [2]$_{5/2}$&111958.89& ($^{3}$P$_{2}$)6p&[2]$_{5/2}$ &  \\
           &     &        &    &$-$0.45  &2&\\
           &     &        &  &$-$0.609N  &3,4&\\
           &     &        &  &\multicolumn{3}{l}{$-$0.562$\pm$0.020ZBD}\\
    6051.15 &1000 & 6051.120&$-$0.03 &$-$0.252N  &1,2,3&   95437.67 &($^{3}$P$_{2}$)5d & [3]$_{7/2}$&111958.89& ($^{3}$P$_{2}$)6p&[2]$_{5/2}$  &  \\
            &     &        &  &$-$0.35    &4\\
            &     &        &  &\multicolumn{3}{l}{$-$0.257$\pm$0.020ZBD}\\
 
   6097.59 & 1000& 6097.57&$-$0.02  &$-$0.45 &1,3&   95396.74 &($^{3}$P$_{2}$)5d & [2]$_{5/2}$&111792.17& ($^{3}$P$_{2}$)6p&[2]$_{3/2}$ & &\\
           &     &         &  &$-$0.35  &2&\\
           &     &         &  &$-$0.30  &4&\\
           &     &        &  &$-$0.237N &\\
           &     &        &  &\multicolumn{3}{l}{$-$0.355$\pm$0.025ZBD}\\ 
   6101.43& 200  &  6101.37&$-$0.06  &$-$0.70 & 1,4& 107904.50 &($^{3}$P$_{1}$)5d & [1]$_{3/2}$&124289.45& ($^{3}$P$_{1}$)6p&[1]$_{3/2}$  &2 noise, 4 blend\\
           &      &       &    &$-$0.10  &3&\\

\hline
\noalign{\smallskip}
\end{tabular}
\end{flushleft}
\end{table*}

\setcounter{table}{0}

\begin{table*}[!hbp]
\caption[ ]{cont.} 
\font\grande=cmr7
\grande
\begin{flushleft}
\begin{tabular}{rrlllrrcllllll}
\hline\noalign{\smallskip}
\multicolumn{1}{c}{$\lambda$(Lab)}&
\multicolumn{1}{c}{Int.}&
\multicolumn{1}{c}{$\lambda$(stellar)}&
\multicolumn{1}{c}{$\Delta\lambda$}&
\multicolumn{1}{c}{$\log~gf$}&
\multicolumn{1}{c}{}&
\multicolumn{1}{c}{$\chi_{low}$(cm$^{-1}$)}&
\multicolumn{2}{c}{Term}&
\multicolumn{1}{c}{$\chi_{up}$(cm$^{-1}$)}&
\multicolumn{2}{c}{Term}&
\multicolumn{1}{l}{Notes}\\
\hline\noalign{\smallskip}

    6194.07&300  &  6194.07&~0.00  &$-$0.10 & 1& 124070.06 &($^{1}$D$_{2}$)5d & [1]$_{3/2}$&140209.99& ($^{3}$P$_{2}$)4f&[2]$_{5/2}$ &  2 noise, 4 blend\\
           &     &         &  &$+$0.20  &3&\\ 
    6270.82&400  &  6270.82&$-$0.01  &$-$0.35 & 1& 112924.84 &($^{1}$D$_{2}$)6s & [2]$_{3/2}$&128867.20& ($^{1}$D$_{2}$)6p&[3]$_{5/2}$ &1 blend, 4 noise\\
           &     &        &   &$-$0.10  &2,3&\\
          &     &        &   &$-$0.196N&\\ 

    6277.54&300  &$--$& & $-$0.894N&   &96033.48 &($^{3}$P$_{2}$)5d & [2]$_{3/2}$&111958.89& ($^{3}$P$_{2}$)6p&[2]$_{5/2}$ & 1,2,3,4 in telluric\\
           &     &  &  &\multicolumn{3}{l}{$-$0.778$\pm$0.021ZBD}\\
    6300.86&100& 6300.830&$-$0.03 & $-$1.10 & 1&        &                  &            &         &                  &            &          2,3,4 noise \\
    6343.96&300  &6343.94&$-$0.02&  $-$0.80 & 1&  96033.48 &($^{3}$P$_{2}$)5d & [2]$_{3/2}$&111792.17& ($^{3}$P$_{2}$)6p&[2]$_{3/2}$ & \\
          &     &       &  &$-$0.65  &2&\\
          &     &6343.95&$-$0.01  &$-$0.55  &3,4&\\
          &     &    &     &\multicolumn{3}{l}{$-$0.786$\pm$0.024ZBD}\\
    6356.35&500  &6356.33&$-$0.02 &$-$0.25 & 1& 124301.96 &($^{1}$D$_{2}$)5d & [2]$_{5/2}$&140029.99& ($^{3}$P$_{2}$)4f&[4]$_{7/2}$ &  2,3,4 noise\\
    6375.28 &100 &6375.28 &$+$0.00&$-$1.00  &1 & 105947.55 &($^{3}$P$_{2}$)5d & [0]$_{1/2}$&121628.82& ($^{3}$P$_{0}$)6p&[1]$_{3/2}$ &  2,3,4 noise\\
    6512.83 &300 &6512.79&$-$0.04 &$-$1.00  &1,2,3 & 107904.50 &($^{3}$P$_{1}$)5d & [1]$_{3/2}$&123254.60& ($^{3}$P$_{1}$)6p&[2]$_{3/2}$ & 4 blend telluric\\
    6528.65 &200 &6528.65 &$+$0.00&$-$0.40  &1 & 114751.08 &($^{3}$P$_{1}$)5d & [3]$_{5/2}$&130063.96& ($^{1}$D$_{2}$)6p&[3]$_{7/2}$ &  2,3,4 noise\\
    6595.01 &800& 6594.97 &$-$0.04&$+$ 0.00  & 1,2,3&   114905.15 &($^{1}$D$_{2}$)5d & [4]$_{9/2}$&130063.96& ($^{1}$D$_{2}$)6p&[3]$_{7/2}$ &
  blend, 4 blend telluric\\
    6597.25 &300& 6597.23 &$-$0.02&$-$0.60&1,2& 106475.21   &($^{3}$P$_{1}$)5d & [1]$_{3/2}$&123254.60& ($^{3}$P$_{1}$)6p&[2]$_{3/2}$  &4 noise\\
    6620.02 & 200&6620.02&$+$0.00 &$-$0.85&1,4&   105313.33 & ($^{3}$P$_{2}$)5d & [1]$_{3/2}$&120414.87& ($^{3}$P$_{0}$)6p&[1]$_{1/2}$  &2,3 noise, 4 blend  \\
    6694.32 & 400&6694.285&$-$0.035&$-$1.00&1,2&    96858.18& ($^{3}$P$_{2}$)5d & [1]$_{1/2}$&111792.17& ($^{3}$P$_{2}$)6p&[2]$_{3/2}$& 4 noise \\
            &    &       &  &$-$0.75  &3\\
            &    &       &  &\multicolumn{3}{l}{$-$0.912$\pm$0.020ZBD}\\
    6788.71 & 100&6788.71&~0.00&$-$0.50&1&  109653.14& ($^{1}$D$_{2}$)6s & [2]$_{5/2}$&124289.45& ($^{3}$P$_{1}$)6p&[1]$_{3/2}$&  2,4 noise\\
    6790.37 & 80& 6790.37 &~0.00& $-$0.70& 1& 106907.120& ($^{3}$P$_{1}$)6s& [1]$_{1/2}$&121628.82 &($^{3}$P$_{0}$)6p& [1]$_{3/2}$&  2 blend, 3,4 noise\\
    6805.74  &1000&$---$ &   &  &  &  108423.070 &($^{3}$P$_{1}$)5d & [3]$_{7/2}$&123112.54& ($^{3}$P$_{1}$)6p&[2]$_{5/2}$ & 1,2,3,4 blend \\
             &    &      &  &$-$0.595N & \\     
             &    &      &  &\multicolumn{3}{l}{$-$0.547$\pm$0.023ZBD}\\
    6990.88 & 2000&6990.835&$-$0.045 & $+$0.25 & 1,2 &   99404.99 &($^{3}$P$_{2}$)5d & [4]$_{9/2}$&113705.40& ($^{3}$P$_{2}$)6p&[3]$_{7/2}$ &\\
            &     &      &  & $+$0.35 & 3,4&\\  
            &     &      &  & $+$0.200N&\\  
            &     &      &  & \multicolumn{3}{l}{$+$0.084$\pm$0.032ZBD}\\
    7082.15 & 200& 7082.15&~0.00 &$+$0.05 &1 & 114751.080 & ($^{3}$P$_{1}$)5d & [3]$_{5/2}$&128867.20& ($^{1}$D$_{2}$)6p&[3]$_{5/2}$  &2,3 noise, 4 blend\\
    7164.83 & 800& 7164.85&$+$0.02 &$+$0.20  & 1,2&  114913.98 &($^{1}$D$_{2}$)5d & [4]$_{7/2}$&128867.20& ($^{1}$D$_{2}$)6p&[3]$_{5/2}$  &3,4 blend telluric\\
    7284.34 & 100& 7284.24&$-$0.10& $-$0.50  &1   & 107904.50  &($^{3}$P$_{1}$)5d & [1]$_{3/2}$&121628.82 &($^{3}$P$_{0}$)6p & [1]$_{3/2}$ & 2,4, 3?? noise \\
    7339.30 & 300& 7339.30 &~0.00&$+$0.45 &1 & 108007.28&($^{3}$P$_{0}$)5d & [2]$_{5/2}$& 121628.82 &($^{3}$P$_{0}$)6p & [1]$_{3/2}$ &  2,4 noise,3?? \\
    7787.04 & 100& 7787.04&~0.00 &$-$0.50?  &1 &119085.49 &($^{1}$D$_{2}$)5d & [5]$_{5/2}$& 131923.79 &($^{1}$D$_{2}$)6p & [5]$_{3/2}$ &  2,3,4 noise  \\
\hline
\noalign{\smallskip}
\end{tabular}
\end{flushleft}
\end{table*}

\end{document}